\documentclass[journal,draftclsnofoot,onecolumn,12pt]{IEEEtran}
\IEEEoverridecommandlockouts
\usepackage{amsmath}
\usepackage{amsfonts}
\usepackage{amssymb}
\usepackage{graphicx}
\usepackage{epstopdf}
\usepackage{caption}
\usepackage{subcaption}
\usepackage{cite}
\usepackage{algorithmicx}
\usepackage[ruled]{algorithm}
\usepackage{algpseudocode}
\usepackage{blindtext}
\usepackage{enumitem}
\usepackage{pdfpages}
\usepackage{bibentry}
\usepackage{amsthm}
\usepackage{bbm}
\usepackage{stackengine}

\begin{document}
%
\title{Throughput--Outage Analysis and Evaluation of Cache-Aided D2D Networks with Measured Popularity Distributions}

\author{Ming-Chun Lee,~\IEEEmembership{Student Member,~IEEE}, Mingyue Ji,~\IEEEmembership{Member,~IEEE}, Andreas F. Molisch,~\IEEEmembership{Fellow,~IEEE}, and Nishanth Sastry
\thanks{M.-C. Lee and A. F. Molisch are with Department of Electrical and Computer Engineering, University of Southern, Los Angeles, CA 90089, USA (email: mingchul@usc.edu, molisch@usc.edu).}
\thanks{M. Ji is with Department of Electrical and Computer Engineering, University of Utah, Salt Lake City, UT 84112, USA (email: mingyue.ji@utah.edu).}
\thanks{N. Sastry is at the Department of Informatics, King's College London, London, UK (e-mail: nishanth.sastry@kcl.ac.uk).}
\thanks{Part of this work will be submitted to the 2019 IEEE Global Communications Conference \cite{MeaPopDist:Lee}.}
}


%


\maketitle

\vspace{-48pt}
\begin{abstract}
Caching of video files on user devices, combined with file exchange through device-to-device (D2D) communications is a promising method for increasing the throughput of wireless networks. Previous theoretical investigations showed that throughput can be increased by orders of magnitude, but assumed a Zipf distribution for modeling the popularity distribution, which was based on observations in {\em wired} networks. Thus the question whether cache-aided D2D video distribution can provide in practice the benefits promised by existing theoretical literature remains open. To answer this question, we provide new results specifically for popularity distributions of video requests of mobile users. Based on an extensive real-world dataset, we adopt a generalized distribution, known as Mandelbrot-Zipf (MZipf) distribution. We first show that this popularity distribution can fit the practical data well. Using this distribution, we analyze the throughput--outage tradeoff of the cache-aided D2D network and show that the scaling law is identical to the case of Zipf popularity distribution when the MZipf distribution is sufficiently skewed, implying that the benefits previously promised in the literature could indeed be realized in practice. To support the theory, practical evaluations using numerical experiments are provided, and show that the cache-aided D2D can outperform the conventional unicasting from base stations.
\end{abstract}


%
\IEEEpeerreviewmaketitle

\vspace{-15pt}
\section{Introduction}

Wireless data traffic is anticipated to increase at a rate of $ 50-100 \% $ per year for the foreseeable future. The main driver for this development is video traffic, which accounts for about $2/3$ of all wireless data  \cite{Cisco:5G}, and has emerged as the ``killer application'' of 4G cellular services, as well as motivating the ``enhanced mobile broadband'' thrust of 5G. It is thus paramount to find cost-effective ways to increase the throughput of cellular networks for video distribution.

Traditional methods for throughput enhancement have treated video traffic just like any other traffic, meaning that each file transmission or on-demand streaming transmission is treated as a unicast. Consequently, it relies on the general throughput enhancement methods of cellular networks such as network densification, HetNets \cite{Andrews_2013}, massive MIMO, and use of additional spectrum (in particular mm-wave bands \cite{Rappaport_2013}). However, these approaches tend to be either very expensive, and/or not scalable.  

On-demand video has, however, unique properties compared to other data, namely (i) high concentration of the popularity distribution (i.e., a small number of videos accounts for the majority of the video traffic), and (ii) {\em asynchronous content reuse}, i.e., those files are watched by different people at different times.\footnote{The latter property distinguishes 
video streaming services such as Netflix, Amazon Prime, Hulu, and Youtube, from the traditional broadcast TV, which achieved high spectral efficiency by forcing viewers to watch particular  videos at prescribed times.} This offers the possibility of employing caching as a part of the video distribution process, thus {\em converting memory into bandwidth} \cite{Ji:Th_Out_toff}, meaning that if we double the memory size of each user device, then the per user throughput can also be doubled. Such an approach is appealing because bandwidth is limited and expensive, while memory is relatively cheap and a rapidly growing hardware resource. Caching approaches include selfish on-device caching \cite{Gol:Dcache1}, femtocaching \cite{Gol:femtocaching}, coded caching \cite{Maddah-Ali:CCache,Ji:order_opt,Maddah-Ali:DCCache,Maddah-Ali:NCCache}, and caching combined with device-to-device (D2D) based file exchange \cite{Gol:Dcache1,Gol:femtocaching,Ji:Th_Out_toff}. On-device caching naturally uses the storage of users' own devices to cache files possibly watched by users in the future. Since this requires predictions of user behavior to gain benefits, designs incorporating recommendation systems are investigated \cite{Karamshuk:PreCache}. Femtocaching or caching in base station (BS) exploits the storage in BSs to relax the requirement of the backhaul \cite{Gol:femtocaching,Bharath:HetCaching}. Coded caching, combining coding with multicasting, leverages the redundancy of cache memories and the broadcasting nature of the wireless medium, and effectively converts memory into bandwidth \cite{Maddah-Ali:CCache,Ji:order_opt,Maddah-Ali:DCCache,Maddah-Ali:NCCache}. Cache-aided D2D exploits recent high-throughput D2D communications \cite{Doppler:D2D} and storage in user devices to gain benefits. This method has been shown to provide not only appealing throughput scaling laws (network throughput increasing linearly with the number of devices) \cite{Ji:Th_Out_toff}, but also robustness in realistic propagation conditions \cite{Ji:Dcache}. It will therefore be the focus of this paper.   

Cache-aided D2D considered in this paper has the following principle: each user device caches, at random, a subset of the video files (the particular caching distribution is a function of the video popularity distribution and other system parameters). When a user requests a file, it might be either already in this user's cache, or is obtained from a nearby device through short-distance D2D communications. This approach was first suggested by one of the authors in \cite{Gol:Asilomar_2011}, and since then has been widely explored in the literature. Among the related papers, different goals, including optimizing outage probability \cite{Ji:Dcache,Ji:Th_Out_toff,Song:D2D-cache,Malak:SpaD2D}, throughput \cite{Ji: Fund_D2D,Ji:Dcache,Ji:Th_Out_toff,Chen:Dcache,Chen:D2D_Coop,Lee:caching}, energy efficiency (EE) \cite{Lee:caching,EE:Chen,Deng:cost}, and delay \cite{Pei:delay,Delay:Wang}, are pursued using various approaches, and different theoretical and practical aspects have been considered. For example, the information-theoretic throughput-outage scaling laws were explored in \cite{Ji:Th_Out_toff, Ji:Dcache,Ji: Fund_D2D}. The tradeoff between throughput and EE was investigated in \cite{Lee:caching}. In \cite{EE:Chen}, battery life was taken into account for optimal EE design. In \cite{Zhang:D2D_Schedule}, a joint scheduling, power control, and caching policy design was proposed. In \cite{Wang:mobil_cach} mobility was leveraged to maximize offloading data to D2D networks. Stochastic geometry was used to analyze cache-aided D2D in \cite{Malak:MaxSucRate,Dhillon:StoCache}. To deal with time-varying popularity distributions, dynamic caching content replacement was discussed in \cite{Lee:Rep_Journal}. Since the field of video caching has been of great interest in the past several years and several hundred related papers have been published, the above literature review cites only a sample of papers and topics.

Most existing papers assume the popularity distribution as the Zipf distribution (essentially a power law distribution). However, this assumption was based on observations in {\em wired} networks \cite{UMass} with Youtube videos and with little empirical support for {\em wireless} network. A recent investigation \cite{Notre-Dame} into wireless popularity distributions of general content showed little content reuse. It is noteworthy that - as the authors of the paper point out - this investigation could not identify video reuse, since video connections were run via a secure https connection, so that the content of the videos could not be determined.\footnote{The paper has been sometimes {\em misinterpreted} as indicating that there is little video reuse.} Therefore, the question remains open whether cache-aided D2D video distribution can achieve the significant gains promised in the literature. This paper aims to answer this question.

In particular, we use the {\em measured} video popularity distribution of the BBC (British Broadcasting Corporation) iPlayer, the most popular video distribution service in the UK. Through appropriate postprocessing, we are able to extract the popularity distribution for the videos watched via cellular connections (these might be different from the files watched through wired connections). We find that this distribution is not well described by a Zipf distribution, but rather a Mandelbrot-Zipf (MZipf) distribution \cite{Hefeeda:P2P}, which is somewhat less skewed.\footnote{A similar result was found in \cite{Maddah-Ali:NCCache}. However, it was not for mobile users.} Such distribution, in contrast to the simple Zipf distribution, is characterized by two parameters: the Zipf factor $\gamma$ and plateau factor $q$, and it degenerates to the Zipf distribution when $q=0$. Thus the MZipf distribution generalizes the Zipf distribution. Considering this more general model, we investigate the benefits of the cache-aided D2D video distribution.

To understand the performance of the cache-aided D2D video distribution, we conduct a thorough throughput--outage tradeoff analysis following the framework in \cite{Ji:Th_Out_toff} but using a different analytical approach and aim to see the scaling law of the throughput-outage tradeoff when the more general MZipf distribution is considered. We derive the analytical formulation of the caching policy maximizing the probability of users to access the desired files via D2D communications. Based on this policy, we obtain the achievable throughput--outage tradeoff. Since the MZipf distribution has the additional factor $q$, the derived caching policy and achievable throughput--outage tradeoff can characterize the influence of $q$. This distinguishes our results from \cite{Ji:Th_Out_toff}. However, this does not imply the resulting scaling behavior is worse than the case with the Zipf distribution. In contrast, the results indicate that, in a practical range of $q$, the same scaling law as considering the Zipf distribution can be obtained again when the MZipf distribution is considered; implying that the benefits promised by existing literature should be retained in practice. We emphasize that, after investigating the real-world data, we find that this range of $q$ is valid in practice.

To support the theoretical analysis, numerical experiments are conducted in D2D networks considering MZipf distributions parameterized based on the real-world data and the realistic setup adopted from \cite{Ji:Dcache}. Results show that the cache-aided D2D scheme can provide orders of magnitude improvement of throughput for a negligible outage probability compared to conventional unicasting. Our main contributions are summarized below:

\begin{itemize}
\item Based on an extensive BBC iPlayer dataset, we extract the popularity distribution for the videos watched by mobile users. Such distribution is then modeled and parameterized by the MZipf distribution, which is a generalized version of the widely used Zipf distribution.
\item To investigate the throughput--outage tradeoff of the cache-aided D2D networks considering a MZipf distribution, we generalize the theoretical treatment of \cite{Ji:Th_Out_toff} with a different but simpler proof technique. 
\item We show that the scaling law of cache-aided D2D achieved in \cite{Ji:Th_Out_toff} is achievable in the case of the practical MZipf distribution; we also characterize the influences of the critical parameters $\gamma$ and $q$ of the MZipf distribution on the throughput--outage tradeoff.
\item To support the theoretical study, we conduct numerical experiments with practical details and show that the cache-aided D2D can significantly outperform the conventional unicasting.
\end{itemize}

The remainder of the paper is organized as follows. In Sec. II, the dataset of video requests for mobile users is described and the corresponding modeling and parameterization are presented. In Sec. III, the theoretical analysis of throughput--outage tradeoff is provided and insights are discussed. We offer numerical experiments in Sec. IV to support the theory. Finally, we conclude the paper in Sec. V. Proofs of theorems and corollaries are relegated to appendices.

Scaling law order notation: given two functions $f$ and $g$, we say that: (1) $f(n)=\mathcal{O}(g(n))$ if there exists a constant $c$ and integer $N$ such that $f(n)\leq cg(n)$ for $n>N$. (2) $f(n)=o(g(n))$ if $\lim_{n\to\infty}\frac{f(n)}{g(n)}=0$. (3) $f(n)=\Omega(g(n))$ if $g(n)=\mathcal{O}(f(n))$. (4) $f(n)=\omega(g(n))$ if $g(n)=o(f(n))$. (5) $f(n)=\Theta(g(n))$ if $f(n)=\mathcal{O}(g(n))$ and $g(n)=\mathcal{O}(f(n))$.

\section{Measured Data and Popularity Distribution Modeling}
This work uses an extensive set of real-world data, namely the dataset of the BBC iPlayer \cite{Karamshuk:PreCache,Nencioni:PreCache,Lee:Indi_pre_model_ToN}, to obtain realistic video demand distributions. The BBC iPlayer is a video streaming service from BBC that provides video content for a number of BBC channels without charge. Content on the BBC iPlayer is available for up to 30 days depending on the policies. We consider two datasets covering June and July, 2014, which include 192,120,311 and 190,500,463 recorded access sessions, respectively. In each record, access information of the video content contains two important columns: \textit{user id} and \textit{content id}. \textit{User id} is based on the long-term cookies that uniquely (in an anonymized way) identify users. \textit{Content id} is the specific identity that uniquely identifies each video content separately. Although there are certain exceptions, \textit{user id} and \textit{content id} can generally help identify the user and the video content of each access. More detailed descriptions of the BBC iPlayer dataset can be found in \cite{Karamshuk:PreCache,Nencioni:PreCache,Lee:Indi_pre_model_ToN}.

To facilitate the investigation, preprocessing is conducted on the dataset. We notice that a user could access the same file multiple times, possibly due to temporary disconnnections from Internet and/or due to temporary pauses by users while moving. Since a user is unlikely to access the same video after finishing watching the video within the period of a month \cite{Lee:Indi_pre_model_ToN}, we consider multiple accesses made by the same user to the same file as a single unique access.

We then separate the data requested by cellular users from those requested via cabled connections or personal WiFi by observing the services of the Internet service providers (ISPs), resulting in $640,631$ different unique accesses (requests) among $267,424$ different users in June; $689,461$ different unique accesses among $327,721$ different users in July. We also separate the data between different regions by observing the Internet gateway through which the requests are routed. For example, for the largest British operator O2, all cellular requests are served via one of three gateways in all of the UK. This on one hand allows an easy separation, but on the other hand does not allow a precise localization of the requests: as outlined below, the whole country can only be divided into 3 regions. Hence, in the following, we will make the assumption that the popularity distribution at each location follows the global (over a particular region) popularity distribution. This tends to underestimate the gains from caching, since it is intuitive that local popularity distributions are more skewed than global popularity distributions. 

\begin{figure}
\mbox{
\hspace{-45pt}
\begin{subfigure}{0.6\textwidth}
\includegraphics[width=\textwidth]{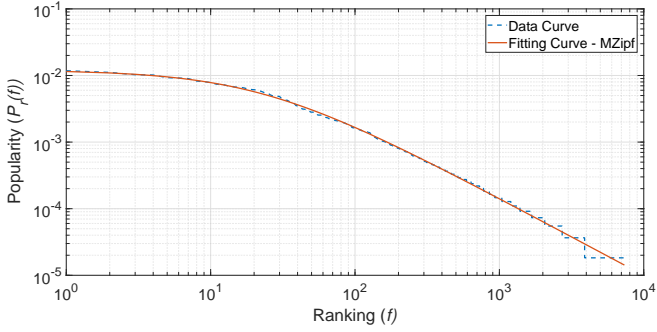}
\caption{Region 2.}
\end{subfigure}
\hspace{-30pt}
\begin{subfigure}{0.6\textwidth}
\includegraphics[width=\textwidth]{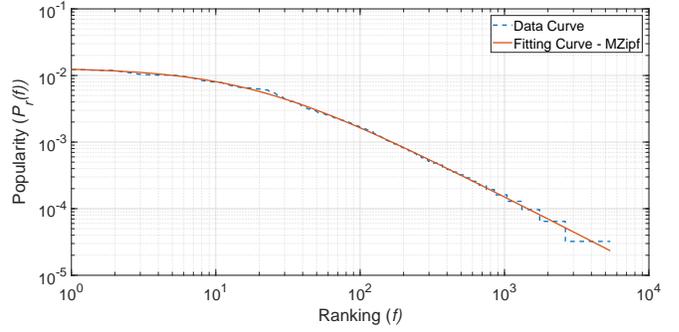}
\caption{Region 3.}
\end{subfigure}
}
\vspace{-10pt}
\caption{Measured ordered popularity distribution of video files of the BBC iPlayer requested via the cellular operator ``O2'' in July of 2014.}
\label{fg:Fig_1}
\vspace{-15pt}
\end{figure}

Based on these data, we plot the global popularity distribution and find that the Zipf distribution is not a good fit. Instead, a MZipf distribution \cite{Hefeeda:P2P} provides a good approximation (see examples in Fig. \ref{fg:Fig_1}):\footnote{Data from other months and regions show similar results. We thus omit their demonstrations for brevity.}
\begin{equation}
\label{eq:gener_pop}
P_r(f)=\frac{(f+q)^{-\gamma}}{\sum_{j=1}^{M} (j+q)^{-\gamma}},f=1,2,...,M,
\end{equation}
where $P_r(f)$ is the probability that users access file $f$,\footnote{We call $P_r(f)$ also the request probability of users for file $f$.} $M$ is the number of files in the library, $\gamma$ is the Zipf factor, and $q$ is the plateau factor. We note that the MZipf distribution degenerates to a Zipf distribution when $q=0$.

\begin{figure}
\center
\includegraphics[width=0.6\textwidth]{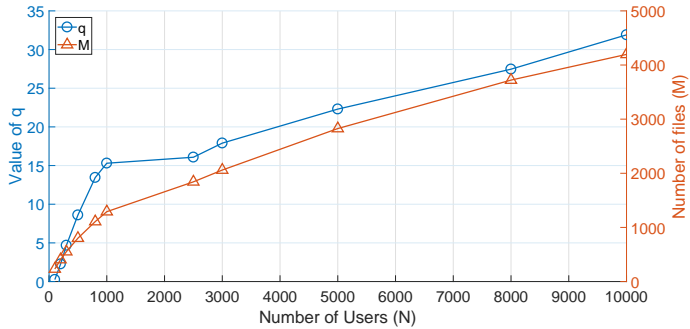}
\vspace{-10pt}
\caption{Relation between $q$, $M$, and $N$ using data in region 1 of June, 2014.}
\vspace{-25pt}
\label{fg:Fig_2}
\end{figure}

A fitting that minimizes the Kullback-Leibler (KL) distance between the data and model provides values of the parameters $\gamma$, $q$, and $M$ as shown in Table \ref{tb:1}.\footnote{The KL distance of a parameter set $\mathbf{x}$ is defined as $D_{\text{KL}}(\mathbf{x})=\sum_m p_m^{\text{data}}\log\frac{p_m^{\text{data}}}{p_m^{\text{model}}(\mathbf{x})}$.} The results imply that up to a breakpoint, i.e., $q$, of approximately 20-50 files, the popularity distribution is relatively flat, and decays faster from there. Also importantly, we find that $\gamma>1$ for all results, which has important implications for the throughput--outage scaling law due to caching. Moreover, we find that the values of $q$ are much smaller (order-wise) than the values of $M$, which also has an important implication that the aggregate memory of the D2D network can easily surpass the number of files requested with similar probabilities and thus should be cached in the D2D network intuitively. Mathematically, in Sec. III, we will see that when the aggregate memory is smaller than the value of $q$ (order-wise), the outage goes to 1 asymptotically as the library size $M$ and $q$ go to infinity, indicating poor performance. Finally, based on the data in region 1 during June, 2014, Fig. \ref{fg:Fig_2} shows the relationship between the values of $q$, $M$, and the number of users $N$; we let $N$ range here from $10$ to $10,000$, covering the range of realistic values for the number of users in a cell. It can be observed that $q$ is much smaller than $M$ when $N$ is realistic. Although not shown here for brevity, $\gamma$ is (on average) between $0.4$ and $1.2$ for the range of $N$ considered in Fig. \ref{fg:Fig_2}, and $\gamma$ generally increases when $N$ increases.

\begin{table}[t]
\caption{Parameterization of Popularity Distribution using the MZipf Model}
\centering
\begin{tabular}{| c | c | c | c | c | c | c |}
\hline
Region & $\gamma$ (June) & $q$ (June) & $M$ (June) & $\gamma$ (July) & $q$ (July) & $M$ (July) \\
\hline
Whole UK & 1.36  & 50  & 16823 & 1.28  & 34  & 19379 \\
\hline
1 & 1.36  & 49  & 16258 & 1.28  & 34  & 18553 \\
\hline
2 & 1.23  & 33  & 6449 & 1.16  & 22  & 7345 \\
\hline
3 & 1.18  & 28  & 4859 & 1.11  & 18  & 5405 \\
\hline
\end{tabular}
\label{tb:1}
\vspace{-20pt}
\end{table} 

\vspace{-10pt}
\section{Achievable Throughput--Outage Tradeoff}
From the measured data, we understand that the MZipf distribution is more suitable for mobile data traffic. In this section, we thus generalize the theoretical treatment in \cite{Ji:Th_Out_toff} by considering the MZipf distribution and provide the achievable throughput--outage tradeoff analysis.
\vspace{-10pt}
\subsection{Network Setup}
In this section, we describe the network model and define the throughput--outage tradeoff. Denote the number of users in the network as $N$. Our goal is to provide the asymptotic analysis when $N\to\infty$, $M\to\infty$, and $q\to\infty$.\footnote{We generally consider $q=\mathcal{O}(M)$ because, by definition, the MZipf distribution would converge to simple uniform distribution when $q=\omega(M)$. Besides, as a matter of practice, we can see from Table \ref{tb:1} that $q$ is much smaller than $M$. Note that we view the case that $q=\Theta(1)$ is a constant simply as a degenerate case of our results. Also, based on the experimental results, $\gamma$ changes within a (small) finite range, i.e., does not go to infinity, as $M$ increases. We therefore approximate $\gamma$ as a fixed constant for the sake of analysis.} We assume a network where user devices can communicate with each other through direct links. We consider the transmission policy using {\em clustering}, in which the devices are grouped geographically into clusters such that any device within one cluster can communicate with any other devices in the same cluster with a constant rate $C$ bits/second/Hz, but not with devices in a different cluster. The network is split into equal-sized clusters. We adopt a grid network in which the users are placed on a regular grid \cite{Ji:Th_Out_toff}. As a result, $g_c(M)\leq N\in\mathbb{N}$, which is a function of $M$ and denoted as the cluster size, is the number of users in a cluster and is a parameter to be chosen in order to analyze the throughput--outage tradeoff. Moreover, we say a potential link exists in the cluster if a user can find its desired file in the cluster through D2D communications and say that a cluster is {\em good} if it contains at least one potential link. 

We assume only a single user in a cluster can use its potential link to obtain the requested file at a time, thus avoiding the interference between users in the same cluster. Besides, potential links of the same cluster are scheduled with equal probability (or, equivalently, in round robin). Therefore all users have the same average throughput. To avoid interference between clusters, we use a spatial reuse scheme with Time Division Multiple Access (TDMA). Denoting $K$ as the reuse factor, such a reuse scheme evenly applies $K$ colors to the clusters,\footnote{We use TDMA only as convenient example. Any scheme that allocates orthogonal resources to clusters with different colors is aligned with our model. } and only the clusters with the same color can be activated on the same time-frequency resource for D2D communications. Note that the adopted reuse scheme is analogous to the spatial reuse scheme in conventional cellular networks \cite{WireCom:Molisch}.

Although the assumptions above are made for the subsequent theoretical analysis, they are actually practical. Specifically, the adjustable size of the cluster can be implemented by adapting the transmit power - in other words, the transmit power is chosen such that communication between opposite corners of a cluster is possible. The link rate for the D2D communication is fixed when no adaptive modulation and coding, and of course this rate has to be smaller than the capacity for the longest-distance communication envisioned in this system. The signal-to-noise ratio (SNR) is determined by the pathloss; small-scale fading can be neglected since in highly frequency-selective channels, the effects of this fading can be eliminated by exploiting the frequency diversity. It must be emphasized that the above network is not optimum for D2D communications. Suitable power control, adaptive modulation and coding, etc., could all increase the spectral efficiency. However, our model provides both a useful lower bound on the performance as well as analytical tractability, which is important for comparability between different schemes. The information theoretical optimal throughput-outage tradeoff analysis is beyond the scope of this paper.

We denote $S$ as the cache memory in a user device, i.e., a user can cache up to $S$ files. Note that we do not consider $S$ to grow to infinity as $N\to\infty$, $M\to\infty$, and $q\to\infty$, i.e., we consider $S=\Theta(1)$ as a fixed network parameter, in this paper. The aggregate memory in a cluster is then $Sg_c(M)$. An independent random caching policy is adopted for users to cache files. Denote $P_c(f)$ as the probability of caching file $f$, where $0\leq P_c(f)\leq 1$ and $\sum_{f=1}^M P_c(f)=1$. Using such caching policy, each user caches each file independently at random according to $P_c(f)$.\footnote{A user might cache the same file multiple times under this caching policy, and this policy is used for the sake of analysis.}

Given the popularity distribution $P_r(\cdot)$, caching policy $P_c(\cdot)$, and transmission policy, we define the average throughput of a user $u$ as $\overline{T}_u=\mathbb{E}\left[ T_{u} \right]$, where $T_{u}$ is a throughput realization of user $u$, and the expectation is taken over the realizations of the cached files and requests.  The minimum average throughput is $\overline{T}_{\text{min}}=\displaystyle{\min_{u}}\overline{T}_u=\overline{T}_u$ due to the symmetry of the network (e.g., round robin scheduling). We define the number of users in outage $N_o$ as the number of users that cannot find their requested files. Thus the average outage is:
\begin{equation}
p_o=\frac{1}{N}\mathbb{E}\left[N_o\right]=\frac{1}{N}\sum_{u}\mathbb{P}\left( \overline{T}_u=0\right)=1-P_u^c,
\end{equation}
where $P_u^c$ is the probability that a user $u$ can find its desired file in a cluster. Due to the symmetry of the network, $P_u^c$ is the same for all users. $P_u^c$ is also called ``hit-rate'' in some literature \cite{Malak:SpaD2D,Chen:Dcache}. We note that our network setup follows the framework in \cite{Ji:Th_Out_toff}. Thus please refer to \cite{Ji:Th_Out_toff} for more rigorous descriptions.

\subsection{Prerequisite for the Analysis of Throughput-Outage Tradeoff}
In this section, we analyze the achievable throughput--outage tradeoff defined by the following:

{\em Definition \cite{Ji:Th_Out_toff}:} For a given network and popularity distribution, a throughput--outage pair $(T, P_o)$ is achievable if there exists a caching policy and a transmission policy with outage probability
$p_o\leq P_o$ and minimum per-user average throughput $\overline{T}_{\text{min}}\geq T$.\footnote{For more comprehensive discussions of the throughput--outage, please see \cite{Ji:Th_Out_toff}.}

Under the network setup considered in Sec. III.A, we determine the throughput--outage tradeoff by adopting the caching policy maximizing $P_u^c$ and by adjusting the cluster size $g_c(M)$. We thus first provide the following theorem:

{\em Theorem 1:} We define $c_2=qa'$, where $a'=\frac{{\gamma}}{S(g_c(M)-1)-1}$, and $c_1\geq 1$ is the solution of the equality $c_1=1+c_2\log\left(1+\frac{c_1}{c_2}\right)$. Let $M\to\infty$, $N\to\infty$, and $q\to\infty$. Suppose $g_c(M)\to\infty$ as $M\to\infty$, and denote $m^*$ as the smallest index such that $P_c^*(m^*+1)=0$. Under the network model in Sec. III.A, the caching distribution $P_c^*(\cdot)$ that maximizes $P_u^c$ is:
\begin{equation}
P_c^*(f)=\left[1-\frac{\nu}{z_f} \right]^+,f=1,...,M,
\end{equation}
where $\nu=\frac{m^*-1}{\sum_{f=1}^{m^*}\frac{1}{z_f}}$, $z_f=\left(P_r(f)\right)^{\frac{1}{S(g_c(M)-1)-1}}$, $[x]^+=\max(x,0)$, and
\begin{equation}
m^*=\Theta\left(\min\left(\frac{c_1Sg_c(M)}{\gamma},M \right)\right).
\end{equation}
\begin{proof}
See Appendix A.
\end{proof}

Observe that $P_c^*(f)$ is monotonically decreasing and $m^*$ determines the number of files whose $P_c^*(f)>0$. Besides, we can observe that $c_1\geq 1$ and $c_1=1$ only if $c_2=o(1)$. Furthermore, we can see that $c_1=\Theta(c_2)$ when $c_2=\Omega(1)$. Thus, when considering $q=\Omega\left(\frac{Sg_c(M)}{\gamma}\right)$ and $\frac{c_1Sg_c(M)}{\gamma}<M$, we obtain $m^*=\Theta(\frac{c_1Sg_c(M)}{\gamma})=\Theta(\frac{c_2Sg_c(M)}{\gamma})=\Theta(q)$. Combining above results, Theorem 1 indicates that the caching policy should cover at least up to the file at rank $q$ (order-wise) in the library. This is intuitive because the MZipf distribution has a relatively flat head and $q$ characterizes the breaking point.

Using the result in Theorem 1, we then characterize $P_u^c$, i.e., the probability that a user can find the desired file in a cluster, in Corollaries 1 and 2:

{\em Corollary 1:} Let $M\to\infty$, $N\to\infty$, and $q\to\infty$. Suppose $g_c(M)\to\infty$ as $M\to\infty$. Consider $q=\mathcal{O}\left(\frac{Sg_c(M)}{\gamma}\right)$ and $g_c(M)<\frac{\gamma M}{c_1 S}$. Under the network model in Sec. III.A and the caching policy in Theorem 1, $P_u^c$ is expressed as:
\begin{equation}
P_u^c=\frac{ \left(\frac{c_1Sg_c(M)}{\gamma}+q\right)^{1-\gamma} }{(M+q)^{1-\gamma}-(q+1)^{1-\gamma}}-\frac{(1-\gamma)\left(\frac{c_1Sg_c(M)}{\gamma}+q\right)^{-\gamma}\left(\frac{c_1Sg_c(M)}{\gamma}\right)}{(M+q)^{1-\gamma}-(q+1)^{1-\gamma}}-\frac{(q+1)^{1-\gamma}}{(M+q)^{1-\gamma}-(q+1)^{1-\gamma}}.
\end{equation}
{\em Corollary 2:} Let $M\to\infty$, $N\to\infty$, and $q\to\infty$. Suppose $g_c(M)\to\infty$ as $M\to\infty$. Consider $q=\mathcal{O}\left(\frac{Sg_c(M)}{\gamma}\right)$ and $g_c(M)=\frac{\rho M}{c_1S}$, where $\rho\geq \gamma$. Define $D=\frac{q}{M}$. Under the network model in Sec. III.A and the caching policy in Theorem 1, $P_u^c$ is lower bounded as
\begin{equation}
P_u^c\geq 1- \frac{(1-\gamma)e^{-(\rho/c_1-\gamma)}}{(1+D)^{1-\gamma}-(D)^{1-\gamma}}\left[(1+D)^{\frac{\gamma}{S(g_c(M)-1)-1}+1}-\left(D\right)^{\frac{\gamma}{S(g_c(M)-1)-1}+1}\right]^{-(S(g_c(M)-1)-1)}.
\end{equation}
\begin{proof}
See Appendix B.
\end{proof}

\subsection{Throughput-Outage Tradeoff for MZipf Distributions with $\gamma<1$}

Using the previous results, we characterize the throughput--outage tradeoff for $\gamma<1$ in the following theorems.

{\em Theorem 2:} Let $M\to\infty$, $N\to\infty$, and $q\to\infty$. Suppose $g_c(M)\to\infty$ as $M\to\infty$. Consider $M=\mathcal{O}(N)$, $q=\mathcal{O}\left(\frac{Sg_c(M)}{\gamma}\right)$, and $\gamma<1$. Denote $\alpha=\frac{1-\gamma}{2-\gamma}$ (i.e.,  $\gamma=\frac{2\alpha-1}{\alpha-1}$). Under the network model in Sec. III.A and the caching policy in Theorem 1, we characterize the throughput--outage tradeoff achievable by adopting the caching policy in Theorem 1 using three regimes:
\begin{enumerate}[label=(\roman*)]
\item Define $c_4=\frac{q}{M^{\alpha}}$. When $g_c(M)=c_3M^{\alpha}$, where $c_3=\Theta(1)$, the achievable throughput--outage tradeoff is
\begin{equation*}
T(P_o)=\frac{C}{K}\frac{M^{-\alpha}}{c_3}\left(1-\exp\left(\frac{-c_3}{2}\left[\left(\frac{Sc_1c_3}{\gamma}+c_4\right)^{{-\gamma}}\left(Sc_1c_3+c_4\right)-(c_4)^{1-\gamma}\right]\right)\right)+o(M^{-\alpha}),
\end{equation*}
where $P_o=1-M^{-\alpha}\left[\left(\frac{Sc_1c_3}{\gamma}+c_4\right)^{{-\gamma}}\left(Sc_1c_3+c_4\right)-(c_4)^{1-\gamma}\right]$.
\item Define $c_5=\frac{q}{g_c(M)}$. When $g_c(M)=\omega(M^{\alpha})<\frac{\gamma M}{c_1S}$, the achievable throughput--outage tradeoff is
\begin{equation*}
T(P_o)=\frac{C}{K}\frac{1}{g_c(M)}+o\left(\frac{1}{g_c(M)}\right),
\end{equation*}
where $P_o=1-\frac{(g_c(M))^{1-\gamma}}{(M+c_5g_c(M))^{1-\gamma}-(c_5g_c(M)+1)^{1-\gamma}}\left[\left(\frac{Sc_1}{\gamma}+c_5\right)^{{-\gamma}}\left(Sc_1+c_5\right)-(c_5)^{1-\gamma}\right]$.
\item Define $D=\frac{q}{M}$. When $g_c(M)=\frac{\rho M}{c_1S}$, where $\rho\geq \gamma$, the achievable throughput-outage tradeoff is
\begin{equation*}
T(P_o)=\frac{C}{K}\frac{Sc_1}{\rho M}+o\left(\frac{1}{M}\right),
\end{equation*}
where $P_o=\frac{(1-\gamma)e^{-(\rho/c_1-\gamma)}}{(1+D)^{1-\gamma}-(D)^{1-\gamma}}\left[(1+D)^{\frac{\gamma}{S(g_c(M)-1)-1}+1}-\left(D\right)^{\frac{\gamma}{S(g_c(M)-1)-1}+1}\right]^{-(S(g_c(M)-1)-1)}$.
\end{enumerate}
\begin{proof}
See Appendix C.
\end{proof}

By comparing Theorem 2 with Theorem 5 in \cite{Ji:Th_Out_toff}, we observe that, when $q=\mathcal{O}\left(\frac{Sg_c(M)}{\gamma}\right)$, the scaling order of the throughput-outage tradeoff in MZipf popularity distribution is identical to that in the Zipf popularity distribution.\footnote{By observing Theorem 2, it is then obvious that we are not interested in cases that $g_c(M)=o(M^{\alpha})$ and $g_c(M)=\omega(M)$ since the former one gives an even worse outage, i.e., $P_o\to 1$, and the latter one gives worse throughput when $P_o\to 0$.}

Theorem 2 indicates that the achievable throughput--outage tradeoff has the same scaling law as the Zipf distribution when the order of $q$ is no larger than the the order of the aggragate memory, indicating that the performance improvement using the cache-aided D2D network with Zipf distribution can be retained when the popularity distribution follows the more practical MZipf distribution. In particular, since other regimes could have unacceptable high outage, the only regime we are interested in is the third regime of Theorem 2. We can then see from the results that the throughput scales with respect to $\Theta(\frac{S}{M})$, meaning that the throughput of cache-aided D2D scales much better than the conventional unicasting when $N$ is much greater than $M$ (small library), i.e., $T\propto \frac{S}{M}>>\frac{1}{N}$.\footnote{Recall that $S$ does not grow to infinity.} Besides, the throughput scales linearly with respect to the memory size of each device. The results also imply that cache-aided D2D has the same scaling law as the coded multicasting scheme of \cite{Maddah-Ali:CCache} and is better than Harmonic Broadcasting \cite{Juhn:HarmBroad}.\footnote{Please refer to \cite{Ji:Th_Out_toff} for detailed discussions.}

Theorem 2 does not characterize the case that $q=\omega\left(\frac{Sg_c(M)}{\gamma}\right)$. We thus provide the relevant discussions. Specifically, we consider the regime that $q=\omega\left(\frac{Sg_c(M)}{\gamma}\right)$ while $q=\mathcal{O}(M)$. This is because when $q=\omega(M)$, the popularity distribution becomes a uniform distribution asymptotically, in which we are not interested. We then provides Theorem 3.

{\em Theorem 3:} Let $M\to\infty$, $N\to\infty$, and $q\to\infty$. Suppose $g_c(M)\to\infty$ as $M\to\infty$. Consider $\gamma<1$, $q=\omega\left(\frac{Sg_c(M)}{\gamma}\right)$, and $q=\mathcal{O}(M)$ (i.e, $g_c(M)=o(M)$). Under the network model in Sec. III.A and the caching policy in Theorem 1, the achievable outage is lower bounded by $1$ asymptotically, i.e., $P_o\geq 1-o(1)$.
\begin{proof}
See Appendix D.
\end{proof}

Theorem 3 suggests that we should increase the cluster size such that the aggregate memory is at least the same order of $q$, i.e., $Sg_c(M)=\Omega(q)$; otherwise the outage will always go to 1. In practice, this implies the outage of the network will be excessive if the aggregate memory is not large enough to accommodate caching at least to the order of $q$ files.

\subsection{Throughput-Outage Tradeoff for MZipf Distributions with $\gamma>1$}
From Theorem 2, we understand that when $\gamma<1$, the only meaningful regime is the third regime. In practice, this implies that it is necessary to have a high density D2D network (or on the other hand, a small library) for realizing the benefits of D2D caching. In this section, we want to see whether this condition can be relaxed when $\gamma>1$, i.e., the popularity is more concentrated on the popular files located in the flat regime of the MZipf distribution. Since Theorem 3 suggests to have a sufficient aggregate memory, we thus focus on the first two regimes of Theorem 2. Specifically, we are interested in the scenario that $g_c(M)=o(M)$ and $q=\mathcal{O}\left(\frac{Sg_c(M)}{\gamma}\right)$:\footnote{We actually can see from Theorem 4 that we need $q=\mathcal{O}\left(\frac{Sg_c(M)}{\gamma}\right)$ to bound $P_o$ away from 1 since $P_o\to 1$ when $c_6=\omega(1)$.}

{\em Theorem 4:} Let $M\to\infty$, $N\to\infty$, and $q\to\infty$. Suppose $g_c(M)\to\infty$ as $M\to\infty$. Consider $\gamma>1$, $g_c(M)=o(M)\leq N$, and $q=\mathcal{O}\left(\frac{Sg_c(M)}{\gamma}\right)$. Define $c_6=\frac{q}{g_c(M)}$. Under the network model in Sec. III.A and the caching policy in Theorem 1, the achievable throughput--outage tradeoff is
\begin{equation}
T(P_o)=\frac{C}{K}\frac{1}{g_c(M)}+o\left(\frac{1}{g_c(M)}\right),
\end{equation}
where $P_o=(c_6)^{\gamma-1}\frac{Sc_1+c_6}{\left(\frac{Sc_1}{\gamma}+c_6\right)^{\gamma}}$.
\begin{proof}
See appendix E.
\end{proof}
If $q=o\left(\frac{Sg_c(M)}{\gamma}\right)$, we obtain $c_1=1$ and $c_6=o(1)$ by definition. We thus have Corollary 3:

{\em Corollary 3:} Let $M\to\infty$, $N\to\infty$, and $q\to\infty$. Suppose $g_c(M)\to\infty$ as $M\to\infty$. Consider $\gamma>1$, $g_c(M)=o(M)\leq N$, and $q=o\left(\frac{Sg_c(M)}{\gamma}\right)$. Under the network model in Sec. III.A and the caching policy in Theorem 1, the achievable throughput--outage tradeoff is
\begin{equation}
T(P_o)=\frac{C}{K}\frac{1}{g_c(M)}+o\left(\frac{1}{g_c(M)}\right),
\end{equation}
where $P_o=o(1)$.

From Theorem 3 and Corollary 3, we observe that when $\gamma>1$, we obtain the scaling law that is better than $\Theta(\frac{S}{M})$ but worse than $\Theta(\frac{S}{q})$. In practice, it implies that when $\gamma>1$ and the aggregate memory is larger than the order of $q$, the improvement of the cache-aided D2D could still be significant even if we have a large library. This relaxes the condition that we need a small library to have significant benefits when $\gamma<1$.

\subsection{Finite-Dimensional Simulations}
\begin{figure}
\mbox{
\hspace{-45pt}
\begin{subfigure}{0.6\textwidth}
\includegraphics[width=\textwidth]{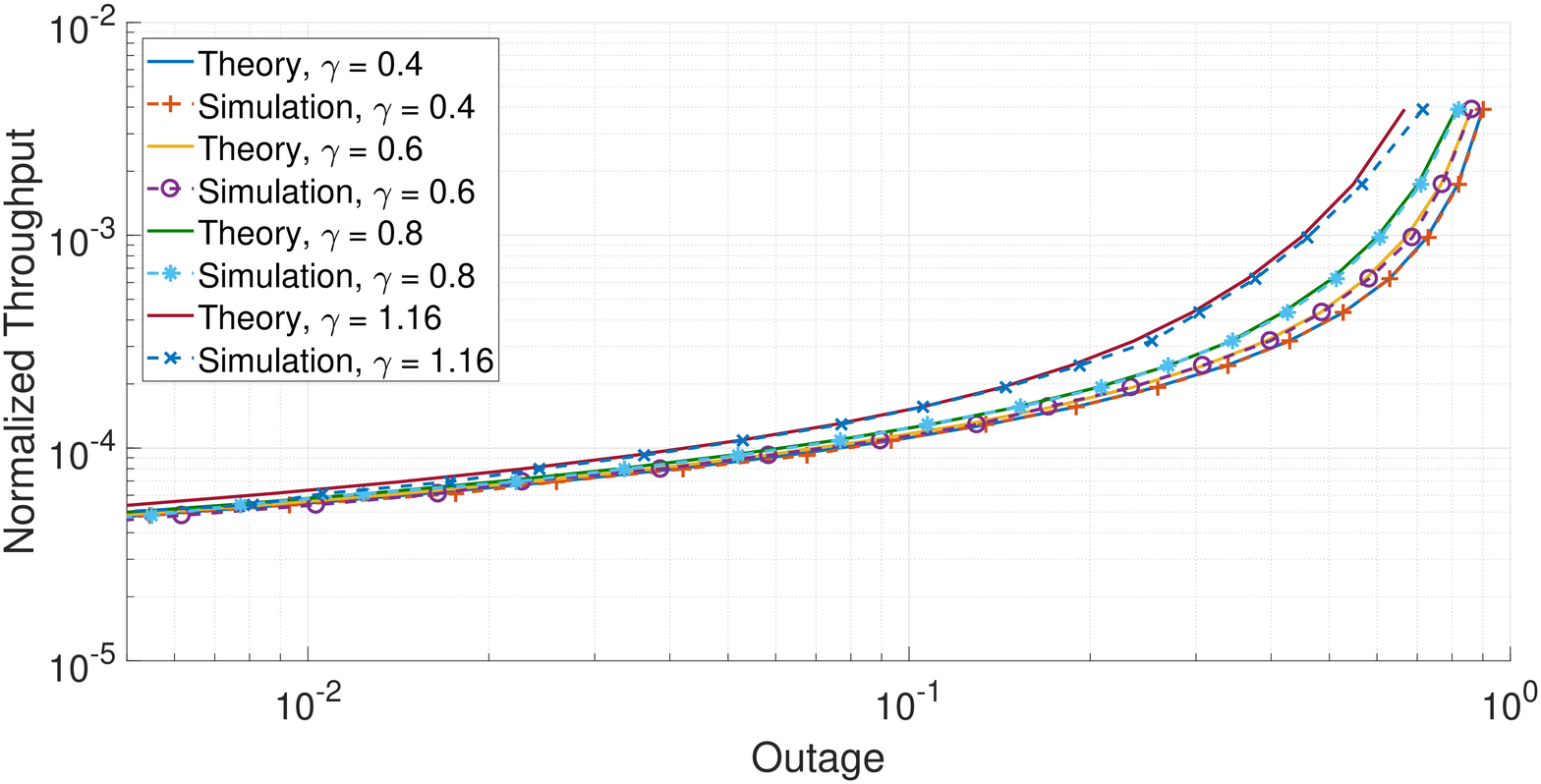}
\caption{Comparisons between different $\gamma$ whose $q=20$.}
\end{subfigure}
\hspace{-30pt}
\begin{subfigure}{0.6\textwidth}
\includegraphics[width=\textwidth]{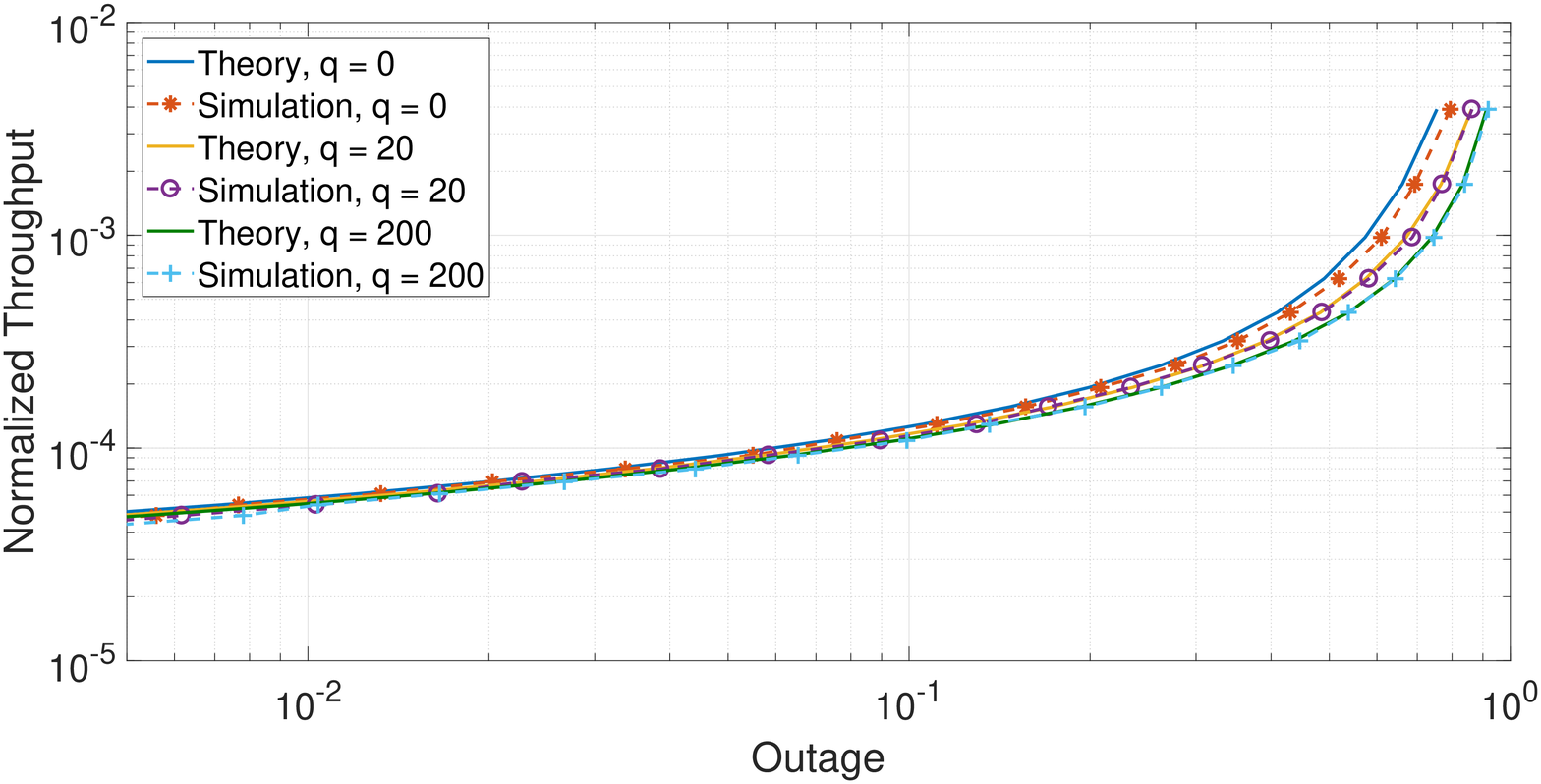}
\caption{Comparisons between different $q$ whose $\gamma=0.6$.}
\end{subfigure}
}
\vspace{-10pt}
\caption{Comparison between the normalized theoretical result (solid lines) and normalized simulated result (dashed lines) in networks adopting $K=4$, $S=1$, $M=1000$, and $N=10000$.}
\label{fg:Fig_3}
\vspace{-20pt}
\end{figure}

Finally, we provide results from finite-dimensional simulations in Fig. \ref{fg:Fig_3}, which compares theoretical (solid lines) and simulated (dashed lines) curves. In Fig. \ref{fg:Fig_3}, we adopt $K=4$, $S=1$, $M=1000$, and $N=10000$. We observe that our analysis can effectively characterize (with small gap) the throughput--outage tradeoff even with finite dimensional setups. This is not common, as indicated by \cite{Ji:Th_Out_toff}, when analyzing the scaling behavior of wireless networks.

\section{Evaluations of cache-aided D2D Networks}
Our theoretical analysis shows that the cache-aided D2D scheme outperforms the conventional unicasting even if the popularity distribution follows the more practical MZipf distribution. To support the theory, we present simulations of the throughput-outage tradeoff using MZipf distributions parameterized according to the real-world data in the network considering practical setups as in \cite{Ji:Dcache}. For the simualtions, communications between users occur at 2.4 GHz. We assume a cell of dimensions  $0.36 {\rm km}^2$ ($600{\rm m} \times 600{\rm m}$) that contains buildings as well as streets/outdoor environments. We assume $N=10000$ users in the cell, i.e., on average, there are $2 \sim 3$ nodes in each  square of $10 \times 10$ meters. The cell contains a Manhattan grid of square buildings with side length of $50$m, separated by streets of width $10$m. Each building is made up of offices of size $6.2 {\rm m} \times 6.2 {\rm m}$. Within the cell, users (devices) are distributed at random according to a uniform distribution. Due to our geometrical setup, each node is assigned to be outdoors or indoors, and in the latter case placed in a particular office. Since $2.4$ GHz communication can penetrate walls, we have to account for different scenarios, which are indoor communication (Winner model A1), outdoor-to-indoor communication (B4), indoor-to-outdoor communication (A2), and outdoor communication (B1) (see \cite{Ji:Dcache}).

The number of clusters in a cell is varied from $2^2=4, 3^2=9,....27^2=729$; a frequency reuse factor of $K=4$ is used to minimize the inter-cluster interference. The cache memory on each device $S$ is kept as a parameter that we will be varied in the simulations. To provide some real-world connections: storage of an hour-long video in medium video quality (suitable for a cellphone) takes about 300 MByte. Thus, storing 100 files with current cellphones is reasonably realistic, and given the continuous increase in memory size, even storing 500 files is not prohibitive (assuming some incentivization by network operators or other entities). 

In terms of channel models, we mostly employ the Winner channel models with some minor modifications. In particular, we directly use Winner II channel models with antenna heights of $1.5$m, as well as the probabilistic Line of Sight (LOS) and Non Line of Sight (NLOS) models. We add a probabilistic body shadowing loss ($\sigma_{L_{b}}$) with a lognormal distribution, where for LOS, $\sigma_{L_{b}} = 4.2$ and for NLOS, $\sigma_{L_{b}} = 3.6$ to account for the blockage of radiation by the person holding the device; see \cite{karedal2008measurement}.\footnote{More details about the channel model can be found in \cite{Ji:Dcache}.}

Since regions 2 and 3 of the dataset cover smaller regions, and thus are expected to describe better the effects that might be encountered within a particular cell (though they are still much larger than a cell), we use their corresponding parameters for MZipf distributions in the simulations.

\begin{figure}
\mbox{
\hspace{-45pt}
\begin{subfigure}{0.6\textwidth}
\includegraphics[width=\textwidth]{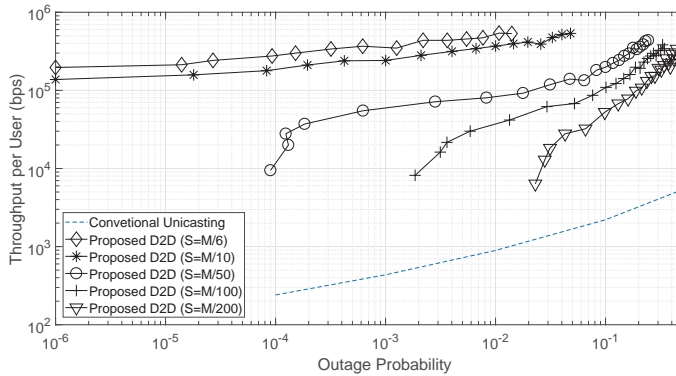}
\caption{Region 2 of July.}
\end{subfigure}
\hspace{-30pt}
\begin{subfigure}{0.6\textwidth}
\includegraphics[width=\textwidth]{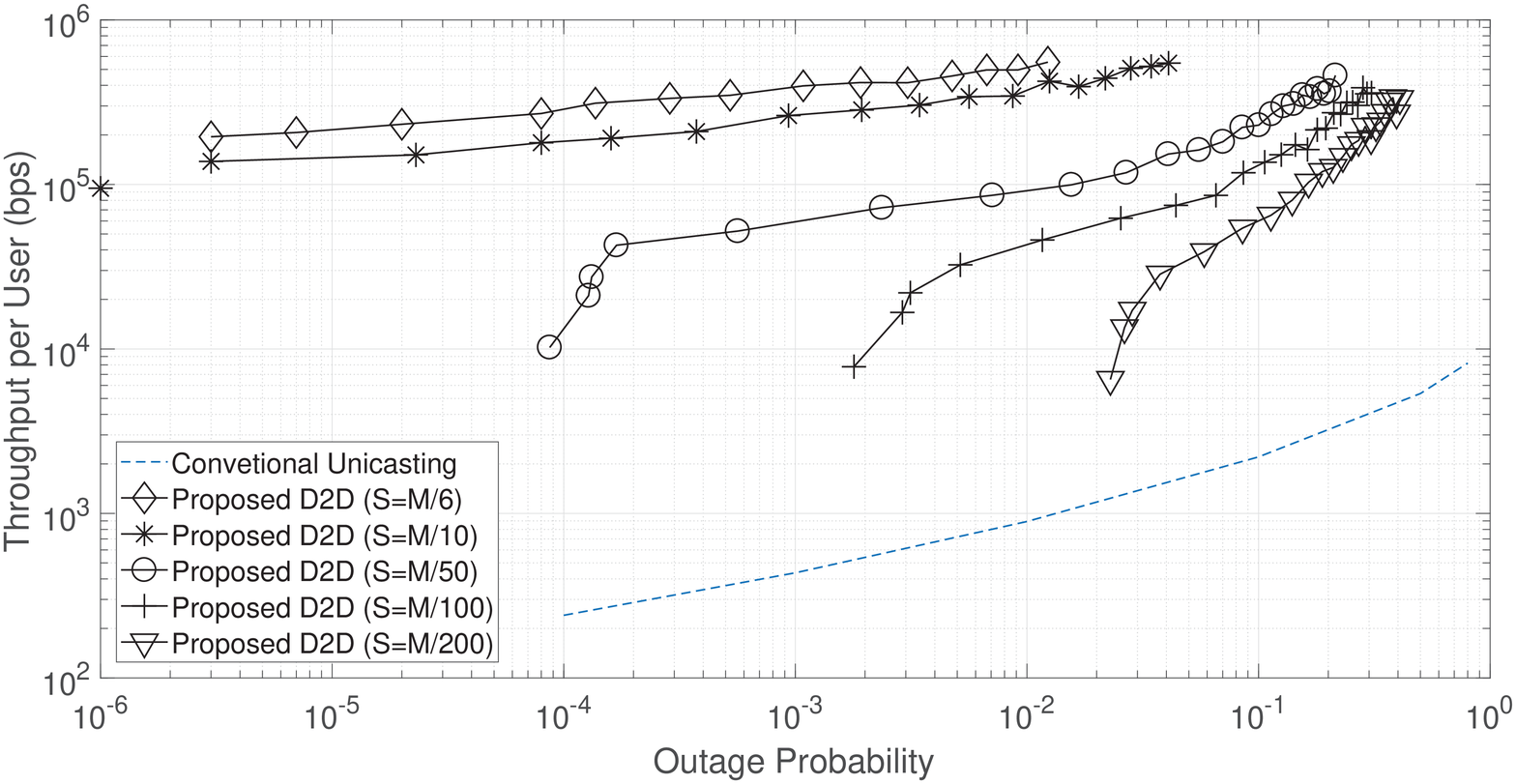}
\caption{Region 3 of July.}
\end{subfigure}
}
\vspace{-10pt}
\caption{Throughput outage tradeoff in networks assuming mixed office scenario for propagation channel; varying local storage size.}
\label{fg:Fig_4}
\vspace{-20pt}
\end{figure}

Fig. \ref{fg:Fig_4}(a) shows the throughput-outage tradeoff for different cache sizes on each device in region 2. An outage of $10\%$ implies that $90 \%$ of traffic can be offloaded to the D2D communications. We can see that the throughput of $10^5$ bps can be achieved if the cache size of each user is up to $1/10$ of the library size. Even for $S=M/50$, i.e., approximate 100 files (30 GB), the advantage compared to conventional unicasting described in \cite{Ji:Dcache} is two orders of magnitude. Even just caching of 30 files ($M/200$) also provides significant throughput gains, though only for outage probabilities $>0.01$.  The results for region 3 (Fig. \ref{fg:Fig_4}(b)) are very similar.

\section{Conclusions}
To answer the open question whether cache-aided D2D for video distribution can provide in practice the benefits promised in the literature, we analyze and evaluate the throughput--outage performance considering measured popularity distributions. Using an extensive dataset, we observe that the widely used Zipf distribution cannot effectively describe the popularity distribution of real wireless traffic data. We thus propose using a generalized version of Zipf distribution, i.e., the MZipf distribution, to model and parameterize the real data. Comparisons with measurements verify the accuracy of this modeling. Considering such generalized modeling, we generalize the theoretical treatment in \cite{Ji:Th_Out_toff} and analyze the throughout--outage tradeoff. In particular, we show the impact of the plateau factor $q$ of the MZipf distribution in the optimal caching distribution and the throughput-outage tradeoff. Theoretical results show that the scaling behavior of the cache-aided D2D is identical to case of Zipf distribution under some parameter regimes validated by real data, implying that the benefits in the case of the Zipf distribution could be retained. To support the theory, extensive numerical evaluations considering practical propagation scenarios and other details are provided, and show that the cache-aided D2D for video distribution significantly outperforms the conventional unicasting. Since the theory and numerical experiments both suggest positive results, we thus conclude that the cache-aided D2D for video distribution can in practice provide the benefits promised in the existing literature.

\appendices
\section{Proof of Theorem 1}
In this section, our goal is to find the caching policy that maximizes $P_u^c$. Note that the probability that a user $u$ can find its desired file $f$ in the cluster through D2D communications is $1-\left(1-P_c(f)\right)^{S(g_c(M)-1)}$. Then by using the law of total probability, we have
$
P_u^c=\sum_{f=1}^M P_r(f)\left(1-\left(1-P_c(f)\right)^{S(g_c(M)-1)}\right).
$
To maximize $P_u^c$, we follow the similar approach based on convex minimization and KKT conditions in Appendix C of \cite{Ji:Th_Out_toff} and obtain 
\begin{equation}
P_c^*(f)=\left[ 1- \left( \frac{\lambda}{P_r(f)S(g_c(M)-1)}\right)^{\frac{1}{S(g_c(M)-1)-1}}  \right]^+.
\end{equation}
Next, we need to find the $\lambda$ such that $\sum_{f=1}^M P_c^*(f) = 1$. Let $\nu=\left( \frac{\lambda}{S(g_c(M)-1)}\right)^{\frac{1}{S(g_c(M)-1)-1}}$ and $z_f=\left(P_r(f)\right)^{\frac{1}{S(g_c(M)-1)-1}}$. Note that $z_f$ is non-increasing with respect to $f$ since $P_r(f)$ is non-increasing. By following the similar argument in appendix C of \cite{Ji:Th_Out_toff}, we obtain that $\nu=\frac{m^*-1}{\sum_{f=1}^{m^*}\frac{1}{z_f}}$, satisfying $\nu\geq z_{m^*+1}$ and $\nu\leq z_{m^*}$. Thus if $m^*$ is a unique integer in $\lbrace 1,2,\cdots,M-1 \rbrace$, it satisfies:
$
m^*\geq 1 + z_{m^*+1}\sum_{f=1}^{m^*}\frac{1}{z_f}
$
and 
$
m^*\leq 1 + z_{m^*}\sum_{f=1}^{m^*}\frac{1}{z_f}.
$
Then in order to determine $m^*$ as a function of $g_c(M)$ in the assumption that $g_c(M)\to \infty$ as $M\to\infty$, we need to evaluate
\begin{equation}\label{eq:low_m_star}
z_{m^*+1}\sum_{f=1}^{m^*}\frac{1}{z_f}=(m^*+q+1)^{\frac{{-\gamma}}{S(g_c(M)-1)-1}}\sum_{f=1}^{m^*}(f+q)^{\frac{{\gamma}}{S(g_c(M)-1)-1}}=\left(\frac{1}{m^*+q+1} \right)^{a'} \sum_{f=1}^{m^*}\left( f+q \right)^{a'},
\end{equation}
\begin{equation}\label{eq:upp_m_star}
z_{m^*}\sum_{f=1}^{m^*}\frac{1}{z_f}=(m^*+q)^{\frac{{-\gamma}}{S(g_c(M)-1)-1}}\sum_{f=1}^{m^*}(f+q)^{\frac{{\gamma}}{S(g_c(M)-1)-1}}=\left(\frac{1}{m^*+q} \right)^{a'} \sum_{f=1}^{m^*}\left( f+q \right)^{a'},
\end{equation}
where $a'=\frac{{\gamma}}{S(g_c(M)-1)-1}$. We then characterize $\sum_{f=1}^{m^*}(f+q)^{a'}$. By using the fundamental concept of integration, we observe that
\begin{equation}\label{eq:upp_int}
\begin{aligned}
&\sum_{f=1}^{m^*}(f+q)^{a'}\leq \int_1^{m^*+1} (x+q)^{a'}dx=\frac{(m^*+q+1)^{a'+1}-(q+1)^{a'+1}}{a'+1},\\
&\sum_{f=1}^{m^*}(f+q)^{a'}\geq (1+q)^{a'} +\int_1^{m^*} (x+q)^{a'}dx
=(1+q)^{a'} +\frac{(m^*+q)^{a'+1}-(q+1)^{a'+1}}{a'+1}.
\end{aligned}
\end{equation}
By using (\ref{eq:upp_int}), we can obtain the upper (UB 1) bound and lower bound (LB 1) of (\ref{eq:low_m_star}):
\begin{equation}
\begin{aligned}
&\text{LB 1}=\left(\frac{q+1}{m^*+q+1}\right)^{a'}+\frac{1}{a'+1}\left[ (m^*+q)\left(\frac{m^*+q}{m^*+q+1}\right)^{a'} - (q+1)\left(\frac{q+1}{m^*+q+1}\right)^{a'} \right],\\
&\text{UB 1}=\frac{1}{a'+1}\left[ (m^*+q+1) - (q+1)\left(\frac{q+1}{m^*+q+1}\right)^{a'} \right].
\end{aligned}
\end{equation}
Similarly, we can obtain the upper (UB 2) bound and lower bound (LB 2) of (\ref{eq:upp_m_star}):
\begin{equation}
\begin{aligned}
&\text{LB 2}=\left(\frac{q+1}{m^*+q}\right)^{a'}+\frac{1}{a'+1}\left[ (m^*+q) - (q+1)\left(\frac{q+1}{m^*+q}\right)^{a'} \right],\\
&\text{UB 2}=\frac{1}{a'+1}\left[ (m^*+q+1)\left(\frac{m^*+q+1}{m^*+q}\right)^{a'} - (q+1)\left(\frac{q+1}{m^*+q}\right)^{a'} \right].
\end{aligned}
\end{equation}
We then define $c_1=m^*a'$ and $c_2=qa'$. Notice that $a'\downarrow 0$ as $g_c(M)\to\infty$. Hence
\begin{equation}
\begin{aligned}\nonumber
\text{LB 1}&=\left(\frac{\frac{c_2}{a'}+1}{\frac{c_1+c_2}{a'}+1}\right)^{a'}+\frac{1}{a'+1}\left[ \left(\frac{c_1+c_2}{a'}\right)\left(\frac{\frac{c_1+c2}{a'}}{\frac{c_1+c_2}{a'}+1}\right)^{a'} - \left(\frac{c_2}{a'}+1\right)\left(\frac{\frac{c_2}{a'}+1}{\frac{c_1+c_2}{a'}+1}\right)^{a'} \right],\\
&=1-\delta_1(a')+\frac{1}{1+a'}\left[ \left(\frac{c_1+c_2}{a'}\right)(1-\delta_2(a')) - \left(\frac{c_2}{a'}+1\right)(1-\delta_1(a')) \right]\\
&=\frac{1}{1+a'}\left[ \left(\frac{c_1+c_2}{a'}\right)(1-\delta_2(a')) - \left(\frac{c_2}{a'}\right)(1-\delta_1(a')) -a'(1-\delta_1(a'))\right]\\
\end{aligned}
\end{equation}
\begin{equation}
\begin{aligned}
\text{UB 1}&=\frac{1}{a'+1}\left[ \left(\frac{c_1+c_2}{a'}+1\right) - \left(\frac{c_2}{a'}+1\right)\left(\frac{\frac{c_2}{a'}+1}{\frac{c_1+c_2}{a'}+1}\right)^{a'} \right]\\
&=\frac{1}{1+a'}\left[ \left(\frac{c_1+c_2}{a'}+1\right) - \left(\frac{c_2}{a'}+1\right)(1-\delta_1(a')) \right],
\end{aligned}
\end{equation}
and
\begin{equation}
\begin{aligned}
\text{LB 2}&=\left(\frac{\frac{c_2}{a'}+1}{\frac{c_1+c_2}{a'}}\right)^{a'}+\frac{1}{a'+1}\left[ \left(\frac{c_1+c_2}{a'}\right) - \left(\frac{c_2}{a'}+1\right)\left(\frac{\frac{c_2}{a'}+1}{\frac{c_1+c_2}{a'}}\right)^{a'} \right]\\
&=\frac{1}{1+a'}\left[ \left(\frac{c_1+c_2}{a'}\right) - \left(\frac{c_2}{a'}\right)(1-\delta_3(a')) -a'(1-\delta_3(a'))\right],\\
\text{UB 2}&=\frac{1}{a'+1}\left[ \left(\frac{c_1+c_2}{a'}+1\right)\left(\frac{\frac{c_1+c_2}{a'}+1}{\frac{c_1+c_2}{a'}}\right)^{a'} - \left(\frac{c_2}{a'}+1\right)\left(\frac{\frac{c_2}{a'}+1}{\frac{c_1+c_2}{a'}}\right)^{a'} \right]\\
&=\frac{1}{1+a'}\left[ \left(\frac{c_1+c_2}{a'}+1\right)(1+\delta_4(a')) - \left(\frac{c_2}{a'}+1\right)(1-\delta_3(a')) \right],
\end{aligned}
\end{equation}
where $\delta_i(a')$, $i=1,...,4$ tend to zeros as $a'\downarrow 0$. Then we denote that 
\begin{equation}
\begin{aligned}
1-\delta_1(a')=\left(\frac{c_2+a'}{c_1+c_2+a'}\right)^{a'}=\left(\nu_1\right)^{a'},1-\delta_2(a')=\left(\frac{c_1+c_2}{c_1+c_2+a'}\right)^{a'}=\left(\nu_2\right)^{a'},\\
1-\delta_3(a')=\left(\frac{c_2+a'}{c_1+c_2}\right)^{a'}=\left(\nu_3\right)^{a'},1-\delta_4(a')=\left(\frac{c_1+c_2+a'}{c_1+c_2}\right)^{a'}=\left(\nu_4\right)^{a'}.\\
\end{aligned}
\end{equation}
It follows that
\begin{equation}
\frac{c}{a'}\delta_i(a')=\frac{c\left[1-\left(\nu_i\right)^{a'}\right]}{a'}\stackrel{a'\to 0}{=}-c\log\left(\nu_i\right),i=1,...,4,
\end{equation}
where the second equality is by L'H\^{o}spital's rule. Thus, suppose $c=\mathcal{O}(c_1+c_2)$, we obtain
\begin{equation}
\begin{aligned}
&\frac{c}{a'}\delta_1(a')\stackrel{a'\to 0}{=}c\log\left(1+\frac{c_1}{c_2}\right),\frac{c}{a'}\delta_2(a')\stackrel{a'\to 0}{=}0,
\frac{c}{a'}\delta_3(a')\stackrel{a'\to 0}{=}c\log\left(1+\frac{c_1}{c_2}\right),\frac{c}{a'}\delta_4(a')\stackrel{a'\to 0}{=}0.
\end{aligned}
\end{equation}
By using the above results and that $m^*=\frac{c_1}{a'}$, it follows that, when $a'\to 0$, we obtain
\begin{equation}
\frac{c_1/a'+\epsilon}{a'+1} + 1 \lessapprox\frac{c_1}{a'}\lessapprox \frac{c_1/a'+\epsilon}{a'+1} + 1,
\end{equation}
where $\epsilon=c_2\log\left(1+\frac{c_1}{c_2}\right)$. Thus, we obtain
\begin{equation}
\frac{c_1}{a'}\left( 1-\frac{1}{a'+1} \right)=\frac{c_1}{a'+1}\cong 1+\frac{\epsilon}{a'+1},
\end{equation}
leading to $c_1\cong a'+1+\epsilon=1+\epsilon.$ We then conclude that
\begin{equation}
m^*=\frac{c_1}{a'}=\frac{c_1(S(g_c(M)-1)-1)}{\gamma}+\mathcal{O}(1),
\end{equation}
where $c_1$ satisfies the equality $c_1=1+c_2\log\left(1+\frac{c_1}{c_2}\right)$ and $c_2=qa'$. This indicates $m^*=\frac{c_1Sg_c(M)}{\gamma}$ to the leading order. Besides, it should be clear that if $\frac{c_1Sg_c(M)}{\gamma}\geq M$, we have $m^*=M$. We also note that when $q=0$, our result degenerates to the results in \cite{Ji:Th_Out_toff} (Observe that when $q=0$, we obtain $c_2=0$ and $c_1=1$).

\section{Proof of Corollary 1 and Corollary 2}
Before starting the main proof, we first provide a useful Lemma:

{\em Lemma 1:} Denote $\displaystyle{\sum_{m=a}^b} (m+q)^{-\gamma}=H(\gamma,q,a,b)$. When $\gamma\neq 1$, we have
\begin{equation*}
\frac{1}{1-\gamma}\left[ (b+q+1)^{1-\gamma}-(a+q)^{1-\gamma}\right]\leq H(\gamma,q,a,b)\leq \frac{1}{1-\gamma}\left[ (b+q)^{1-\gamma}-(a+q)^{1-\gamma} \right] + (a+q)^{-\gamma}.
\end{equation*}
{\em Proof. }Consider $\gamma\neq 1$. By the fundamental calculus, we have
\begin{equation*}
\begin{aligned}
H(\gamma,q,a,b)&=\sum_{m=a}^b (m+q)^{-\gamma}\geq\int_a^{b+1}\frac{dx}{(x+q)^{\gamma}}\\
&=\frac{1}{1-\gamma}(x+q)^{1-\gamma}\mid_a^{b+1}=\frac{1}{1-\gamma}\left[ (b+q+1)^{1-\gamma}-(a+q)^{1-\gamma}\right],\\
H(\gamma,q,a,b)&=\sum_{m=a}^b (m+q)^{-\gamma}\leq (a+q)^{-\gamma} + \int_a^{b}\frac{dx}{(x+q)^{\gamma}}\\
&=(a+q)^{-\gamma} + \frac{1}{1-\gamma}(x+q)^{1-\gamma}\mid_a^{b}=\frac{1}{1-\gamma}\left[ (b+q)^{1-\gamma}-(a+q)^{1-\gamma}\right] + (a+q)^{-\gamma}.
\end{aligned}
\end{equation*}

\subsection{Proof of Corollary 1}
We consider $g_c(M)<\frac{\gamma M}{c_1S}$ and $q=\mathcal{O}\left(\frac{Sg_c(M)}{\gamma}\right)$. We thus obtain $c_1=\mathcal{O}(1)$ and $m^*<M$. The probability that a user $u$ finds the desired file in the cluster is then
\begin{equation}
\begin{aligned}
P_u^c&=\sum_{f=1}^M P_r(f)\left(1-\left(1-P_c(f)\right)^{S(g_c(M)-1)}\right)=\sum_{f=1}^{m^*}P_r(f)\left(1-\left(\frac{\nu}{z_f}\right)^{S(g_c(M)-1)}\right)\\
&\stackrel{(a)}{\leq}\sum_{f=1}^{m^*}P_r(f)-\sum_{f=1}^{m^*}P_r(f)\left(\frac{P_r(m^*+1)}{P_r(f)}\right)\cdot \left(\frac{P_r(m^*+1)}{P_r(f)}\right)^{\frac{1}{S(g_c(M)-1)-1}}\\
&=\sum_{f=1}^{m^*}P_r(f)-P_r(m^*+1)\sum_{f=1}^{m^*} \left(\frac{f+q}{m^*+1+q}\right)^{\frac{\gamma}{S(g_c(M)-1)-1}}\\
&=\frac{H(\gamma,q,1,m^*)}{H(\gamma,q,1,m)}-\frac{(m^*+q+1)^{-\gamma}}{H(\gamma,q,1,m)}\sum_{f=1}^{m^*} \left(\frac{f+q}{m^*+1+q}\right)^{\frac{\gamma}{S(g_c(M)-1)-1}}\\
&\stackrel{(b)}{\leq}\frac{ \frac{1}{1-\gamma}(m^*+q)^{1-\gamma}-\frac{1}{1-\gamma}(q+1)^{1-\gamma} + (1+q)^{-\gamma}- (m^*+q+1)^{-\gamma} \sum_{f=1}^{m^*} \left(\frac{f+q}{m^*+1+q}\right)^{\frac{\gamma}{S(g_c(M)-1)-1}}}{\frac{1}{1-\gamma}(M+q+1)^{1-\gamma}-\frac{1}{1-\gamma}(q+1)^{1-\gamma}}\\
&=\frac{ (m^*+q)^{1-\gamma}-(q+1)^{1-\gamma} + (1-\gamma)(1+q)^{-\gamma}}{(M+q+1)^{1-\gamma}-(q+1)^{1-\gamma}}\\
&\qquad\qquad-\frac{(1-\gamma)(m^*+q+1)^{-\gamma}\left(\frac{1}{m^*+1+q}\right)^{\frac{\gamma}{S(g_c(M)-1)-1}} \sum_{f=1}^{m^*} \left(f+q\right)^{\frac{\gamma}{S(g_c(M)-1)-1}}}{(M+q+1)^{1-\gamma}-(q+1)^{1-\gamma}}\\
&\stackrel{(c)}{\leq}\frac{ (m^*+q)^{1-\gamma}-(q+1)^{1-\gamma} + (1-\gamma)(1+q)^{-\gamma}}{(M+q+1)^{1-\gamma}-(q+1)^{1-\gamma}}\\
&-\frac{(1-\gamma)(m^*+q+1)^{-\gamma}\left(\frac{1}{m^*+1+q}\right)^{\frac{\gamma}{S(g_c(M)-1)-1}} \left[ \left(1+q\right)^{\frac{\gamma}{S(g_c(M)-1)-1}}+\int_1^{m^*}(x+q)^{\frac{\gamma}{S(g_c(M)-1)-1}}dx\right]}{(M+q+1)^{1-\gamma}-(q+1)^{1-\gamma}}\\
&=\frac{ (m^*+q)^{1-\gamma}-(q+1)^{1-\gamma} + (1-\gamma)(1+q)^{-\gamma}}{(M+q+1)^{1-\gamma}-(q+1)^{1-\gamma}}-\frac{(1-\gamma)(m^*+q+1)^{-\gamma}\left(\frac{1}{m^*+1+q}\right)^{\frac{\gamma}{S(g_c(M)-1)-1}}}{(M+q+1)^{1-\gamma}-(q+1)^{1-\gamma}}\\
&\qquad\cdot\left[ \left(1+q\right)^{\frac{\gamma}{S(g_c(M)-1)-1}}+\frac{1}{{\frac{\gamma}{S(g_c(M)-1)-1}}+1}\left( (m^*+q)^{\frac{\gamma}{S(g_c(M)-1)-1}+1}-(q+1)^{{\frac{\gamma}{S(g_c(M)-1)-1}}+1} \right)\right]\\
&\stackrel{(d)}{=}\frac{ \left(\frac{c_1Sg_c(M)}{\gamma}+q\right)^{1-\gamma}-(q+1)^{1-\gamma}}{(M+q)^{1-\gamma}-(q+1)^{1-\gamma}}-\frac{(1-\gamma)\left(\frac{c_1Sg_c(M)}{\gamma}+q\right)^{-\gamma}\left(\frac{c_1Sg_c(M)}{\gamma}\right)}{(M+q)^{1-\gamma}-(q+1)^{1-\gamma}}\\
&\qquad\qquad+o\left(\frac{\left(\frac{c_1Sg_c(M)}{\gamma}+q\right)^{1-\gamma}}{(M+q)^{1-\gamma}-(q+1)^{1-\gamma}}\right),
\end{aligned}
\end{equation}
where $(a)$ is because $\nu\geq z_{m^*+1}$; $(b)$ uses results in Lemma 1; $(c)$ exploits Riemann sum and $m^*\geq 1$; $(d)$ uses Theorem 1 that $m^*=\frac{c_1Sg_c(M)}{\gamma}$ and $g_c(M)\to\infty$. Similarly,
\begin{equation*}
\begin{aligned}
P_u^c&\stackrel{(a)}{\geq}\sum_{f=1}^{m^*}P_r(f)-\sum_{f=1}^{m^*}P_r(f)\left(\frac{P_r(m^*)}{P_r(f)}\right)\cdot \left(\frac{P_r(m^*)}{P_r(f)}\right)^{\frac{1}{S(g_c(M)-1)-1}}\\
&=\frac{H(\gamma,q,1,m^*)}{H(\gamma,q,1,m)}-\frac{(m^*+q)^{-\gamma}}{H(\gamma,q,1,m)}\sum_{f=1}^{m^*} \left(\frac{f+q}{m^*+q}\right)^{\frac{\gamma}{S(g_c(M)-1)-1}}\\
&\stackrel{}{\geq}\frac{ \frac{1}{1-\gamma}(m^*+q+1)^{1-\gamma}-\frac{1}{1-\gamma}(q+1)^{1-\gamma}- (m^*+q)^{-\gamma} \sum_{f=1}^{m^*} \left(\frac{f+q}{m^*+q}\right)^{\frac{\gamma}{S(g_c(M)-1)-1}}}{\frac{1}{1-\gamma}(M+q)^{1-\gamma}-\frac{1}{1-\gamma}(q+1)^{1-\gamma}+ (1+q)^{-\gamma}}\\
\end{aligned}
\end{equation*}
\begin{equation}
\begin{aligned}
&=\frac{ (m^*+q+1)^{1-\gamma}-(q+1)^{1-\gamma} }{(M+q)^{1-\gamma}-(q+1)^{1-\gamma}+ (1-\gamma)(1+q)^{-\gamma}}\\
&\qquad\qquad-\frac{(1-\gamma)(m^*+q)^{-\gamma}\left(\frac{1}{m^*+q}\right)^{\frac{\gamma}{S(g_c(M)-1)-1}} \sum_{f=1}^{m^*} \left(f+q\right)^{\frac{\gamma}{S(g_c(M)-1)-1}}}{(M+q)^{1-\gamma}-(q+1)^{1-\gamma}+ (1-\gamma)(1+q)^{-\gamma}}\\
&\stackrel{}{\geq}\frac{ (m^*+q+1)^{1-\gamma}-(q+1)^{1-\gamma}}{(M+q)^{1-\gamma}-(q+1)^{1-\gamma}+ (1-\gamma)(1+q)^{-\gamma}}\\
&\qquad\qquad-\frac{(1-\gamma)(m^*+q)^{-\gamma}\left(\frac{1}{m^*+q}\right)^{\frac{\gamma}{S(g_c(M)-1)-1}} \left[\int_1^{m^*+1}(x+q)^{\frac{\gamma}{S(g_c(M)-1)-1}}dx\right]}{(M+q)^{1-\gamma}-(q+1)^{1-\gamma}+ (1-\gamma)(1+q)^{-\gamma}}\\
&\stackrel{}{=}\frac{ \left(\frac{c_1Sg_c(M)}{\gamma}+q\right)^{1-\gamma}-(q+1)^{1-\gamma} + }{(M+q)^{1-\gamma}-(q+1)^{1-\gamma}}-\frac{(1-\gamma)\left(\frac{c_1Sg_c(M)}{\gamma}+q\right)^{-\gamma}\left(\frac{Sg_c(M)}{\gamma}\right)}{(M+q)^{1-\gamma}-(q+1)^{1-\gamma}}\\
&+o\left(\frac{\left(\frac{c_1Sg_c(M)}{\gamma}+q\right)^{1-\gamma}}{(M+q)^{1-\gamma}-(q+1)^{1-\gamma}}\right),
\end{aligned}
\end{equation}
where $(a)$ is because $\nu\leq z_{m^*}$. By combining the above results, Corollary 1 is proved.

\subsection{Proof of Corollary 2}
When $g_c(M)=\frac{\rho M}{c_1S}$, where $\rho\geq \gamma$, we obtain $m^*=M$. Thus, results in Corollary 1 is no longer appropriate. Now since $m^*=M$, we thus have $\nu=\frac{M-1}{\sum_{f=1}^M\frac{1}{z_f}}$. We define $D=\frac{q}{M}$. Then
\begin{equation*}
\begin{aligned}
P_u^c&=\sum_{f=1}^{M}P_r(f)\left(1-\left(\frac{\nu}{z_f}\right)^{S(g_c(M)-1)}\right)=1-\nu^{S(g_c(M)-1)}\sum_{f=1}^M\frac{P_r(f)}{(z_f)^{S(g_c(M)-1)}}\\
&=1-\left(\frac{M-1}{\sum_{f=1}^M P_r(f)^{\frac{-1}{S(g_c(M)-1)-1}}}\right)^{S(g_c(M)-1)}\sum_{f=1}^MP_r(f)^{\frac{-1}{S(g_c(M)-1)-1}}\\
\end{aligned}
\end{equation*}
\begin{equation*}
\begin{aligned}
&=1-\left(M-1\right)^{S(g_c(M)-1)}\left(\sum_{f=1}^M \left(\frac{(f+q)^{-\gamma}}{H(\gamma,q,1,M)}\right)^{\frac{-1}{S(g_c(M)-1)-1}}\right)^{-(S(g_c(M)-1)-1)}\\
&=1-\frac{\left(M-1\right)^{S(g_c(M)-1)}}{H(\gamma,q,1,M)}\cdot\frac{1}{\left(\sum_{f=1}^M \left(f+q\right)^{\frac{\gamma}{S(g_c(M)-1)-1}}\right)^{(S(g_c(M)-1)-1)}}\\
\end{aligned}
\end{equation*}
Denoting $S(g_c(M)-1)-1$ as $\varphi$, we have
\begin{equation*}
\begin{aligned}
P_u^c&\stackrel{}{\geq}1-\frac{\left(M-1\right)^{S(g_c(M)-1)}}{\frac{1}{1-\gamma}(M+q+1)^{1-\gamma}-\frac{1}{1-\gamma}(q+1)^{1-\gamma}}\cdot\frac{1}{\left((1+q)^{\frac{\gamma}{\varphi}}+\int_1^M(x+q)^{\frac{\gamma}{\varphi}}dx\right)^{\varphi}}\\
&=1-\frac{(1-\gamma)\left(M-1\right)^{S(g_c(M)-1)}}{(M+q+1)^{1-\gamma}-(q+1)^{1-\gamma}}\frac{1}{\left[(1+q)^{\frac{\gamma}{\varphi}}+\left(\frac{1}{\frac{\gamma}{\varphi}+1}\right)\left((M+q)^{\frac{\gamma}{\varphi}+1}-(q+1)^{\frac{\gamma}{\varphi}+1}\right)\right]^{\varphi}}\\
&=1-\frac{(1-\gamma)\left(M-1\right)^{S(g_c(M)-1)}}{(M+q+1)^{1-\gamma}-(q+1)^{1-\gamma}}\cdot\underbrace{\left(1-\frac{\gamma}{\varphi+\gamma}\right)^{(\varphi+\gamma)\frac{\varphi}{\varphi+\gamma}(-1)}}_{=e^{\gamma}}\\
&\cdot\left[\left(\frac{\gamma}{\varphi}+1\right)(1+q)^{\frac{\gamma}{\varphi}}+(M+q)^{\frac{\gamma}{\varphi}+1}-(q+1)^{\frac{\gamma}{\varphi}+1}\right]^{-\varphi}\\
&\stackrel{}{=}1-(1-\gamma)\frac{\left(M-1\right)^{S(g_c(M)-1)}}{\left(M\right)^{S(g_c(M)-1)}}\frac{M^{1-\gamma}}{(M+DM+1)^{1-\gamma}-(DM+1)^{1-\gamma}}\cdot e^{\gamma}\\
&\cdot \Bigg[\left(\frac{\gamma}{\varphi}+1\right)\frac{1}{M}
\cdot\left(\frac{1+DM}{M}\right)^{\frac{\gamma}{\varphi}}+(1+D)^{\frac{\gamma}{\varphi}+1}-\left(D+\frac{1}{M}\right)^{\frac{\gamma}{\varphi}+1}\Bigg]^{-\varphi}\\
&=1-(1-\gamma)\underbrace{\left(1-\frac{1}{M}\right)^{S(\frac{\rho M}{c_1S}-1)}}_{=e^{-\rho/c_1}}\frac{1}{(1+D+\frac{1}{M})^{1-\gamma}-(D+\frac{1}{M})^{1-\gamma}}\cdot e^{\gamma}\\
&\cdot \Bigg[\left(\frac{\gamma}{\varphi}+1\right)\frac{1}{M}\left(D+\frac{1}{M}\right)^{\frac{\gamma}{\varphi}}+(1+D)^{\frac{\gamma}{\varphi}+1}-\left(D+\frac{1}{M}\right)^{\frac{\gamma}{\varphi}+1}\Bigg]^{-\varphi}\\
&=1- \frac{(1-\gamma)e^{-(\rho/c_1-\gamma)}}{(1+D)^{1-\gamma}-(D)^{1-\gamma}}\left[(1+D)^{\frac{\gamma}{\varphi}+1}-\left(D\right)^{\frac{\gamma}{\varphi}+1}\right]^{-\varphi}+o(1).\\
\end{aligned}
\end{equation*}

\section{Proof of Theorem 2}
In this section, we provide the proof for Theorem 2, which lets $M\to\infty$, $N\to\infty$, and $q\to\infty$ and consider $g_c(M)\to\infty$ as $M\to\infty$. We first outline the proof. From Corollaries 1 and 2, we obtain the lower bound of $P_u^c$, which is determined by the cluster size $g_c(M)$ and the condition of $q$. Then since the outage probability $P_o=1-P_u^c$, we can obtain the upper bound of the outage. Subsequently, for each outage regime, we obtain the lower bound of $\overline{T}_{\text{min}}$ by computing the lower bound of the sum throughput $\overline{T}_{\text{sum}}$ and using the result that $\overline{T}_{\text{min}}=\frac{1}{N}\overline{T}_{\text{sum}}$, following the fact that each user is symmetric and has the same average throughput. Since the achievable upper bound of the outage probability and the corresponding lower bound of the throughput can be obtained, we characterize the achievable throughput-outage tradeoff. In Theorem 2, we consider $\gamma<1$ and the regime covering $q=\mathcal{O}\left(\frac{Sg_c(M)}{\gamma}\right)$.\footnote{We will see that only this regime gives the acceptable outage performance when $\gamma<1$.} The cases that $\gamma<1$ and $q=\omega\left(\frac{Sg_c(M)}{\gamma}\right)$ will be considered later in Theorem 3.

The main flow for computing $\overline{T}_{\text{sum}}$ is the following (see also Appendix D in \cite{Ji:Th_Out_toff}). Denote $L$ as the number of active links, we have
\begin{equation}\label{eq:T_sum}
\overline{T}_{\text{sum}}=C\cdot \mathbb{E}[L]=C\cdot E[\text{number of active cluster}],
\end{equation}
where $C$ is the constant link rate and the second equality is because only one transmission is allowed in a cluster in a time-frequency slot. Then noticing that 
\begin{equation}\label{eq:lowbound_Tmin}
\begin{aligned}
\mathbb{E}[\text{number of active cluster}]&\geq \frac{1}{K}\mathbb{E}[\text{number of good cluster}]\\
&=\frac{1}{K}\left(\text{number of total clusters}\cdot \mathbb{P}(W>0)\right),
\end{aligned}
\end{equation}
where the $K$ is reuse factor. Recall that a good cluster is where there exists at least one potential link in the cluster. Thus $W=\sum_{u=1}^{g_c(M)}\mathbf{1}_u$ is the number of potential links, where $\mathbf{1}_u$ is the indicator that equals to one if user $u$ can access the desired file in the cluster; otherwise $\mathbf{1}_u=0$.

\subsection{Proof of Regime 1}
In this section, we consider $g_c(M)=c_3M^{\alpha}$, where $c_3=\Theta(1)$, and $q=\mathcal{O}\left(\frac{Sg_c(M)}{\gamma}\right)$. We define $c_4=\frac{q}{M^{\alpha}}$. According to Corollary 1, we obtain:

\begin{equation*}
\begin{aligned}
P_u^c&=\frac{\left(\frac{c_1Sg_c(M)}{\gamma}+q\right)^{{1-\gamma}}}{(M+q)^{1-\gamma}-(q+1)^{1-\gamma}}-\frac{(1-\gamma)\frac{c_1Sg_c(M)}{\gamma}\left(\frac{c_1Sg_c(M)}{\gamma}+q\right)^{{-\gamma}}}{(M+q)^{1-\gamma}-(q+1)^{1-\gamma}}-\frac{\left(q+1\right)^{{1-\gamma}}}{(M+q)^{1-\gamma}-(q+1)^{1-\gamma}}\\
&=\frac{\left(\frac{c_1c_3SM^{\alpha}}{\gamma}+c_4M^{\alpha}\right)^{{1-\gamma}}}{(M+q)^{1-\gamma}-(q+1)^{1-\gamma}}-\frac{(1-\gamma)\frac{c_1c_3SM^{\alpha}}{\gamma}\left(\frac{c_1c_3SM^{\alpha}}{\gamma}+c_4M^{\alpha}\right)^{{-\gamma}}}{(M+q)^{1-\gamma}-(q+1)^{1-\gamma}}-\frac{\left(c_4M^{\alpha}+1\right)^{{1-\gamma}}}{(M+q)^{1-\gamma}-(q+1)^{1-\gamma}}\\
&\stackrel{(a)}{=}\frac{M^{\alpha(1-\gamma)}}{M^{1-\gamma}}\left[\left(\frac{Sc_1c_3}{\gamma}+c_4\right)^{{1-\gamma}}-(1-\gamma)\frac{Sc_1c_3}{\gamma}\left(\frac{Sc_1c_3}{\gamma}+c_4\right)^{{-\gamma}}-(c_4)^{1-\gamma}\right]+o\left(\frac{M^{\alpha(1-\gamma)}}{M^{1-\gamma}}\right)\\
\end{aligned}
\end{equation*}
\begin{equation*}
\begin{aligned}
&\stackrel{}{=}M^{(\alpha-1)(1-\gamma)}\left[\left(\frac{Sc_1c_3}{\gamma}+c_4\right)^{{-\gamma}}\left(Sc_1c_3+c_4\right)-(c_4)^{1-\gamma}\right]+o\left(M^{-\alpha}\right)\\
&\stackrel{(b)}{=}M^{-\alpha}\left[\left(\frac{Sc_1c_3}{\gamma}+c_4\right)^{{-\gamma}}\left(Sc_1c_3+c_4\right)-(c_4)^{1-\gamma}\right]+o\left(M^{-\alpha}\right),
\end{aligned}
\end{equation*}
where $(a)$ is because $q=o(M)$ and $M\to\infty$, and $(b)$ is because
\begin{equation}
(\alpha-1)(1-\gamma)\stackrel{(d)}{=}(\alpha-1)\left(1-\frac{2\alpha-1}{\alpha-1}\right)=(\alpha-1)\left(\frac{-\alpha}{\alpha-1}\right)=-\alpha,
\end{equation}
where $(d)$ is because
\begin{equation}\label{eq:qqq}
\alpha=\frac{1-\gamma}{2-\gamma}=>2\alpha-\alpha\gamma=1-\gamma=>(\alpha-1)\gamma=2\alpha-1=>\gamma=\frac{2\alpha-1}{\alpha-1}.
\end{equation}

Now we lower bound $\overline{T}_{\text{min}}$ by using (\ref{eq:lowbound_Tmin}) and computing $\mathbb{P}(W>0)$.\footnote{Note that the proof technique used for this part is based on the concentration of functions with the self-bounding property and is different from the one in \cite{Ji:Th_Out_toff}.} We first introduce the definition of self-bounding property and a corresponding Lemma:

{\em Definition \cite{Ji:order_opt}:} Let $\mathcal{X}\subseteq \mathbb{R}$ and consider a non-negative $\nu$-variate function $g:\mathcal{X}\rightarrow[0,\infty)$. We say that $g$ has the self-bounding property if there exists a function $g_i:\mathcal{X}^{\nu-1}\rightarrow\mathbb{R}$ such that, for all ${x_1,...,x_{\nu}}\subseteq\mathcal{X}^{\nu}$ and all $i=1,...,\nu$,
\begin{equation}
\begin{aligned}
&0\leq g_i(x_1,\cdots,x_{\nu})-g_i(x_1,\cdots,x_{i-1},x_{i+1},\cdots,x_{\nu})\leq 1,\\
&\sum_{i=1}^{\nu})(g_i(x_1,\cdots,x_{\nu})-g_i(x_1,\cdots,x_{i-1},x_{i+1},\cdots,x_{\nu}))\leq g(x_1,\cdots,x_{\nu}).
\end{aligned}
\end{equation}

{\em Lemma 2 (p. 182, Th. 6.12 in \cite{Boucheron:inequalities}):} Consider $\mathcal{X}\subseteq \mathbb{R}$ and the random vector $X =(X_1,...,X_\nu )\in \mathcal{X}^\nu$, where $X_1,...,X_{\nu}$ are mutually statistically independent. Denote $Y=g(X)$, where $g(.)$ has the self-bounding property. Then, for any $0<\nu\leq \mathbb{E}[Y]$, we have
\begin{equation}
\mathbb{P}(Y-\mathbb{E}[Y]\leq-\mu)\leq\exp\left(-\frac{\mu^2}{2\mathbb{E}[Y]}\right).
\end{equation}

We observe that the sum function $g(x_1,...,x_{\nu})=\sum_{i=1}^{\nu}x_i$ has self-bounding property when $x_i,\forall i,$ are binary, i.e., $x_i \in \lbrace 0,1\rbrace$. Thus, $W=\sum_{u=1}^{g_c(M)}\mathbf{1}_u$ satisfies the conditions of Lemma 2. By using Lemma 2 and considering $\mu=\mathbb{E}[W]$, we obtain $\mathbb{P}(W\leq 0)\leq \exp\left(-\frac{\mathbb{E}[W]}{2}\right)$. It follows that
\begin{equation}\label{eq:W_ineq}
\mathbb{P}(W> 0)>1-\exp\left(-\frac{\mathbb{E}[W]}{2}\right).
\end{equation}
Using (\ref{eq:lowbound_Tmin}) and (\ref{eq:W_ineq}), we thus obtain
\begin{equation}\label{eq:T_min_1}
\begin{aligned}
&\mathbb{E}[\text{number of active cluster}]\geq \frac{1}{K}\left(\text{number of total clusters}\cdot \mathbb{P}(W>0)\right)\\
&\geq\frac{N}{K g_c(M)}\exp\left(1-\exp\left(-\frac{\mathbb{E}[W]}{2}\right)\right)=\frac{N}{K c_3M^{\alpha}}\left(1-\exp\left(-\frac{\mathbb{E}[W]}{2}\right)\right).
\end{aligned}
\end{equation}
To compute $\mathbb{E}[W]$, we note that $\mathbb{E}[W]=E\left[\sum_{u=1}^{g_c(M)}\mathbf{1}_u\right]=g_c(M)P_u^c$. Thus,
\begin{equation}\label{eq:W_min_eq_1}
\begin{aligned}
\mathbb{E}[W]&=c_3M^{\alpha}\left(M^{-\alpha}\left[\left(\frac{Sc_1c_3}{\gamma}+c_4\right)^{{-\gamma}}\left(Sc_1c_3+c_4\right)-(c_4)^{1-\gamma}\right]+o\left(M^{-\alpha}\right)\right)\\
&=c_3\left[\left(\frac{Sc_1c_3}{\gamma}+c_4\right)^{{-\gamma}}\left(Sc_1c_3+c_4\right)-(c_4)^{1-\gamma}\right]+o(1).
\end{aligned}
\end{equation}
By using (\ref{eq:T_sum}), (\ref{eq:T_min_1}), (\ref{eq:W_min_eq_1}), and that $\overline{T}_{\text{min}}=\frac{1}{N}\overline{T}_{\text{sum}}$, we obtain
\begin{equation*}
\overline{T}_{\text{min}}\geq \frac{C}{K}\frac{M^{-\alpha}}{c_3}\left(1-\exp\left(\frac{-c_3}{2}\left[\left(\frac{Sc_1c_3}{\gamma}+c_4\right)^{{-\gamma}}\left(Sc_1c_3+c_4\right)-(c_4)^{1-\gamma}\right]\right)\right)+o(M^{-\alpha}).
\end{equation*}
Finally, by exploiting the perturbation argument similar to appendix J in \cite{Ji:Th_Out_toff}, we obtain the achievable throughput-outage tradeoff for regime 1 in the theorem as
\begin{equation}
T(P_o)=\frac{C}{K}\frac{M^{-\alpha}}{c_3}\left(1-\exp\left(\frac{-c_3}{2}\left[\left(\frac{Sc_1c_3}{\gamma}+c_4\right)^{{-\gamma}}\left(Sc_1c_3+c_4\right)-(c_4)^{1-\gamma}\right]\right)\right)+o(M^{-\alpha}),
\end{equation}
where $P_o=1-M^{-\alpha}\left[\left(\frac{Sc_1c_3}{\gamma}+c_4\right)^{{-\gamma}}\left(Sc_1c_3+c_4\right)-(c_4)^{1-\gamma}\right]$.

\subsection{Proof of Regime 2}
In this section, we consider $g_c(M)=\omega(M^{\alpha})<\frac{\gamma M}{c_1S}$ and $q=\mathcal{O}\left(\frac{Sg_c(M)}{\gamma}\right)$. We define $c_5=\frac{q}{g_c(M)}$. Again by using Corollary 1, we obtain

\begin{equation*}
\begin{aligned}
P_u^c&=\frac{\left(\frac{c_1Sg_c(M)}{\gamma}+q\right)^{{1-\gamma}}}{(M+q)^{1-\gamma}-(q+1)^{1-\gamma}}-\frac{(1-\gamma)\frac{c_1Sg_c(M)}{\gamma}\left(\frac{c_1Sg_c(M)}{\gamma}+q\right)^{{-\gamma}}}{(M+q)^{1-\gamma}-(q+1)^{1-\gamma}}-\frac{\left(q+1\right)^{{1-\gamma}}}{(M+q)^{1-\gamma}-(q+1)^{1-\gamma}}\\
&=\frac{\left(\frac{c_1Sg_c(M)}{\gamma}+c_5g_c(M)\right)^{{1-\gamma}}-(1-\gamma)\frac{c_1Sg_c(M)}{\gamma}\left(\frac{c_1Sg_c(M)}{\gamma}+c_5g_c(M)\right)^{{-\gamma}}-\left(c_5g_c(M)+1\right)^{{1-\gamma}}}{(M+c_5g_c(M))^{1-\gamma}-(c_5g_c(M)+1)^{1-\gamma}}\\
\end{aligned}
\end{equation*}
\begin{equation}
\begin{aligned}
&=\frac{(g_c(M))^{1-\gamma}\left[\left(\frac{Sc_1}{\gamma}+c_5\right)^{{1-\gamma}}-(1-\gamma)\frac{Sc_1}{\gamma}\left(\frac{Sc_1}{\gamma}+c_5\right)^{{-\gamma}}-(c_5)^{1-\gamma}\right]}{(M+c_5g_c(M))^{1-\gamma}-(c_5g_c(M)+1)^{1-\gamma}}\\
&=\frac{(g_c(M))^{1-\gamma}\left[\left(\frac{Sc_1}{\gamma}+c_5\right)^{{-\gamma}}\left(Sc_1+c_5\right)-(c_5)^{1-\gamma}\right]+o\left(\left(\frac{g_c(M)}{M}\right)^{1-\gamma}\right)}{(M+c_5g_c(M))^{1-\gamma}-(c_5g_c(M)+1)^{1-\gamma}}.
\end{aligned}
\end{equation}
Then we again use the same approach as used in regime 1 to obtain the lower bound of $\overline{T}_{\text{min}}$. We first compute 
\begin{equation}
\begin{aligned}\label{eq:W_min_eq_2}
&E[W]=g_c(M)P_u^c\\
&=\frac{g_c(M)(g_c(M))^{1-\gamma}\left[\left(\frac{Sc_1}{\gamma}+c_5\right)^{{-\gamma}}\left(Sc_1+c_5\right)-(c_5)^{1-\gamma}\right]}{(M+c_5g_c(M))^{1-\gamma}-(c_5g_c(M)+1)^{1-\gamma}}+o\left(\frac{(g_c(M))^{2-\gamma}}{M^{1-\gamma}}\right)\stackrel{(a)}{=} \infty,
\end{aligned}
\end{equation}
where $(a)$ is because $g_c(M)<\frac{\gamma M}{c_1S}$, $q=\mathcal{O}\left(\frac{Sg_c(M)}{\gamma}\right)$, and
\begin{equation}
g_c(M)\left(\frac{g_c(M)}{M}\right)^{1-\gamma}\stackrel{(b)}{=}g_c(M)\omega\left(\frac{M^{\alpha(1-\gamma)}}{M^{1-\gamma}}\right)\stackrel{(c)}{=}g_c(M)\omega(M^{-\alpha})\stackrel{(d)}{=}\omega(1)=\infty
\end{equation}
where $(b)$ is because $g_c(M)=\omega(M^{\alpha})$; $(c)$ follows the same derivations as in (\ref{eq:qqq}); $(d)$ is again because $g_c(M)=\omega(M^{\alpha})$. Consequently, we obtain
\begin{equation}
\overline{T}_{\text{min}}\geq \frac{C}{K}\frac{1}{g_c(M)}+o\left(\frac{1}{g_c(M)}\right),
\end{equation}
since $\exp(-E[W]/2)\to 0$. Again by using a perturbation argument, it follows that
\begin{equation}
T(P_o)=\frac{C}{K}\frac{1}{g_c(M)}+o\left(\frac{1}{g_c(M)}\right),
\end{equation}
where $P_o=1-\frac{(g_c(M))^{1-\gamma}}{(M+c_5g_c(M))^{1-\gamma}-(c_5g_c(M)+1)^{1-\gamma}}\left[\left(\frac{Sc_1}{\gamma}+c_5\right)^{{-\gamma}}\left(Sc_1+c_5\right)-(c_5)^{1-\gamma}\right]$.

\subsection{Proof of Regime 3}
Finally, we consider $g_c(M)=\frac{\rho M}{c_1S}$, where $\rho\geq \gamma$ and $q=\mathcal{O}\left(\frac{Sg_c(M)}{\gamma}\right)$. Thus, instead of using Corollary 1, Corollary 2 is adopted. By Corollary 2, we obtain
\begin{equation}
P_o\leq\frac{(1-\gamma)e^{-(\rho/c_1-\gamma)}}{(1+D)^{1-\gamma}-(D)^{1-\gamma}}\left[(1+D)^{\frac{\gamma}{S(g_c(M)-1)-1}+1}-\left(D\right)^{\frac{\gamma}{S(g_c(M)-1)-1}+1}\right]^{-(S(g_c(M)-1)-1)}+o(1).
\end{equation}
To compute the lower bound of $\overline{T}_{\text{min}}$, it is clear that $\mathbb{E}[W]\to\infty$ because both $P_u^c$ and $g_c(M)$ in regime 3 are larger than their counterparts in regime 2. Consequently, we obtain
\begin{equation}
\begin{aligned}
\overline{T}_{\text{min}}&\geq \frac{C}{K}\frac{1}{g_c(M)}+o\left(\frac{1}{g_c(M)}\right)=\frac{C}{K}\frac{Sc_1}{\rho M}+o\left(\frac{1}{g_c(M)}\right).
\end{aligned}
\end{equation}
Again by using a perturbation argument, we obtain the achievable throughput-outage tradeoff:
\begin{equation}
T(P_o)=\frac{C}{K}\frac{Sc_1}{\rho M}+o\left(\frac{1}{M}\right),
\end{equation}
where $P_o=\frac{(1-\gamma)e^{-(\rho/c_1-\gamma)}}{(1+D)^{1-\gamma}-(D)^{1-\gamma}}\left[(1+D)^{\frac{\gamma}{S(g_c(M)-1)-1}+1}-\left(D\right)^{\frac{\gamma}{S(g_c(M)-1)-1}+1}\right]^{-(S(g_c(M)-1)-1)}$.

\section{Proof of Theorem 3}
Observe that $P_o$ goes to 1 when we consider $q=\mathcal{O}\left(\frac{Sg_c(M)}{\gamma}\right)$ and $g_c(M)=o(M)<\frac{\gamma M}{c_1S}$ according to Theorem 2 (regimes 1 and 2). By intuition, it follows that $P_o$ also goes to 1 when we consider $q=\omega\left(\frac{Sg_c(M)}{\gamma}\right)$ while $q=\mathcal{O}(M)$ since increasing the value of $q$ degrades the concentration of the popularity distribution which increases the outage. This leads to Theorem 3. Rigorously, observe that
\begin{equation}
P_u^c=\sum_{f=1}^M P_r(f)\left(1-\left(1-P_c(f)\right)^{S(g_c(M)-1)}\right)=\sum_{f=1}^M P_r(f)G(f),
\end{equation} 
where $G(f)=\left(1-\left(1-P_c(f)\right)^{S(g_c(M)-1)}\right)$. Then denote the optimal caching policy for $P_r(f;\gamma,q_1)$ as $P_{c}^{q_1}(f)$ and the optimal caching policy for $P_r(f;\gamma,q_2)$ as $P_{c}^{q_2}(f)$, where both $P_{c}^{q_1}(f)$ and $P_{c}^{q_2}(f)$ are monotonically decreasing with respect to $f$ (see Appendix A). Considering $q_1<q_2$, we want to show the following
\begin{equation}\label{eq:bbb}
\begin{aligned}
&\sum_{f=1}^M P_r(f;\gamma,q_1)\left(1-\left(1-P_{c}^{q_1}(f)\right)^{S(g_c(M)-1)}\right)\stackrel{(a)}{\geq}\\
& \sum_{f=1}^M P_r(f;\gamma,q_1)\left(1-\left(1-P_{c}^{q_2}(f)\right)^{S(g_c(M)-1)}\right)\stackrel{(b)}{>}\sum_{f=1}^M P_r(f;\gamma,q_2)\left(1-\left(1-P_{c}^{q_2}(f)\right)^{S(g_c(M)-1)}\right),
\end{aligned}
\end{equation}
is true. Since $(a)$ is true simply because $P_{c}^{q_1}(f)$ is the optimal policy for $P_r(f;\gamma,q_1)$, it is thus sufficient to show that $(b)$ is true.

To show the $(b)$ of (\ref{eq:bbb}) is true, we note that when $g<h$ and $\epsilon>0$,
\begin{equation}\label{eq:abc}
\sum_{f=1}^{M} P_r(f)G(f) + \epsilon G(g) -\epsilon G(h)>\sum_{f=1}^{M} P_r(f)G(f)
\end{equation}
because $G(f)$ is monotonically decreasing when $P_c(f)$ is monotonically decreasing with respect to $f$. Eq. (\ref{eq:abc}) indicates that, given the caching policy is monotonically decreasing, when we add $\epsilon$ to the popularity with lower index (better rank) by subtracting $\epsilon$ from the one with higher index, we can improve $P_u^c$. Then notice that when $q_1<q_2$, we obtain:
\begin{equation}\label{eq:kkk}
P_r(1;\gamma,q_1)=\frac{(1+q_1)^{-\gamma}}{\sum_{f=1}^M (f+q_1)^{-\gamma}}\geq\frac{(1+q_2)^{-\gamma}}{\sum_{f=1}^M (f+q_2)^{-\gamma}}=P_r(1;\gamma,q_2)
\end{equation} 
and
\begin{equation}\label{eq:aaa}
\frac{(f+q_1)^{-\gamma}}{(f+1+q_1)^{-\gamma}}>\frac{(f+q_2)^{-\gamma}}{(f+1+q_2)^{-\gamma}},f=1,2,...,M,
\end{equation}
i.e., starting with a larger value, $P_r(f;\gamma,q_1)$ decreases faster than $P_r(f;\gamma,q_2)$ with respect to $f$. By using (\ref{eq:abc}), (\ref{eq:kkk}), and (\ref{eq:aaa}), we can then obtain
\begin{equation*}
\sum_{f=1}^M P_r(f;\gamma,q_1)\left(1-\left(1-P_{c}^{q_2}(f)\right)^{S(g_c(M)-1)}\right)-\sum_{f=1}^M P_r(f;\gamma,q_2)\left(1-\left(1-P_{c}^{q_2}(f)\right)^{S(g_c(M)-1)}\right)>0,
\end{equation*}
proving the $(b)$ of (\ref{eq:bbb}) is true.
\vspace{-5pt}
\section{Proof of Theorem 4}
We consider $g_c(M)=o(M)\leq N$ and $q=\mathcal{O}\left(\frac{Sg_c(M)}{\gamma}\right)$. Since these regimes imply $g_c(M)<\frac{\gamma M}{c_1S}$, we should apply Corollary 1. We define $c_6=\frac{q}{g_c(M)}$. When $\gamma>1$, we obtain
\begin{equation*}
\begin{aligned}
P_u^c&=\frac{\left(\frac{c_1Sg_c(M)}{\gamma}+q\right)^{{1-\gamma}}}{(M+q)^{1-\gamma}-(q+1)^{1-\gamma}}-\frac{(1-\gamma)\frac{c_1Sg_c(M)}{\gamma}\left(\frac{c_1Sg_c(M)}{\gamma}+q\right)^{{-\gamma}}}{(M+q)^{1-\gamma}-(q+1)^{1-\gamma}}-\frac{\left(q+1\right)^{{1-\gamma}}}{(M+q)^{1-\gamma}-(q+1)^{1-\gamma}}\\
&=\frac{\left(c_6g_c(M)+1\right)^{{1-\gamma}}}{(c_6g_c(M)+1)^{1-\gamma}-(M+c_6g_c(M))^{1-\gamma}}-\frac{\left(\frac{c_1Sg_c(M)}{\gamma}+c_6g_c(M)\right)^{{1-\gamma}}}{(c_6g_c(M)+1)^{1-\gamma}-(M+c_6g_c(M))^{1-\gamma}}\\
&\qquad-\frac{(\gamma-1)\frac{c_1Sg_c(M)}{\gamma}\left(\frac{c_1Sg_c(M)}{\gamma}+c_6g_c(M)\right)^{{-\gamma}}}{(c_6g_c(M)+1)^{1-\gamma}-(M+c_6g_c(M))^{1-\gamma}}\\
&\stackrel{(a)}{\geq}\frac{\left(c_6g_c(M)+1\right)^{{1-\gamma}}}{(c_6g_c(M)+1)^{1-\gamma}}-\frac{\left(\frac{c_1Sg_c(M)}{\gamma}+c_6g_c(M)\right)^{{1-\gamma}}}{(c_6g_c(M)+1)^{1-\gamma}}
-\frac{(\gamma-1)\frac{c_1Sg_c(M)}{\gamma}\left(\frac{c_1Sg_c(M)}{\gamma}+c_6g_c(M)\right)^{{-\gamma}}}{(c_6g_c(M)+1)^{1-\gamma}}\\
&=1-\left(\frac{c_6g_c(M)+1}{\frac{c_1Sg_c(M)}{\gamma}+c_6g_c(M)}\right)^{{\gamma-1}}
-\frac{(\gamma-1)(c_6g_c(M)+1)^{\gamma-1}}{\left(\frac{c_1Sg_c(M)}{\gamma}+c_6g_c(M)\right)^{{\gamma}}\left(\frac{c_1Sg_c(M)}{\gamma}\right)^{-1}}\\
&=1-\left(\frac{c_6g_c(M)+1}{\frac{c_1Sg_c(M)}{\gamma}+c_6g_c(M)}\right)^{{\gamma}}\left(\left(\frac{c_6g_c(M)+1}{\frac{c_1Sg_c(M)}{\gamma}+c_6g_c(M)}\right)^{-1}+(\gamma-1)\left(\frac{c_6g_c(M)+1}{\frac{c_1Sg_c(M)}{{\gamma}}}\right)^{-1}\right)\\
&=1-\left(\frac{c_6g_c(M)+1}{\frac{c_1Sg_c(M)}{\gamma}+c_6g_c(M)}\right)^{{\gamma}}\frac{c_6g_c(M)+c_1Sg_c(M)}{c_6g_c(M)+1}\\
&=1-\left(\frac{c_6}{\frac{Sc_1}{\gamma}+c_6}\right)^{{\gamma}}\frac{c_6+Sc_1}{c_6}-o(1)=1-(c_6)^{\gamma-1}\frac{Sc_1+c_6}{\left(\frac{Sc_1}{\gamma}+c_6\right)^{\gamma}}-o(1),
\end{aligned}
\end{equation*}
where $(a)$ is because $(1+c_6g_c(M))^{1-\gamma}>(1+c_6g_c(M))^{1-\gamma}-(M+c_6g_c(M))^{1-\gamma}>0$. Then notice that $\mathbb{E}[W]=g_c(M)P_u^c\to\infty$ since $g_c(M)\to\infty$ and $c_6=\mathcal{O}(1)$. Consequently, $P(W>0)\to 1$ by Lemma 2 (see Appendix C.A). It follows that
\begin{equation*}
\overline{T}_{\text{min}}\geq \frac{C}{K}\frac{1}{g_c(M)}+o\left(\frac{1}{g_c(M)}\right).
\end{equation*}
Finally, by the perturbation argument again, we obtain Theorem 4.
\vspace{-5pt}


\bibliographystyle{IEEEtran}
%
\nobibliography{IEEEabrv}

\end{document}